# Explainable Deep Neural Network for Multimodal ECG Signals: Intermediate vs Late Fusion

**Timothy Oladunni[1], Ehimen Aneni[2]**

**[1]Computer Science Department, Morgan State University and Yale University**
**Timothy.oladunni@morgan.edu**

**[2]Section of Cardiovascular Medicine, Department of Internal Medicine, Yale University School of Medicine**
**ehimen.aneni@yale.edu**

**This work was supported in part by the ASCEND Research Fellowship, Yale University**

**ABSTRACT** The limitations of unimodal deep learning models, particularly their tendency to overfit and limited generalizability, have renewed interest in multimodal fusion strategies. The multimodal deep neural network (MDNN) has the capability of integrating diverse data domains. However, the optimal fusion strategy, intermediate fusion (feature-level) versus late fusion (decision-level), remains insufficiently examined. We believe this is important in high-stakes clinical contexts such as ECG-based cardiovascular disease (CVD) classification. This study investigates the comparative effectiveness of intermediate and late fusion strategies using ECG signals. Time, frequency, and time-frequency domains were considered. We employed various techniques to identify the most effective fusion architecture. Results demonstrate that intermediate fusion consistently outperformed late fusion, achieving a peak accuracy of 97%, with Cohen's d > 0.8 relative to standalone models and d = 0.40 compared to late fusion. Interpretability analyses using saliency maps reveal that both models align with the discretized ECG signals. Statistical dependency between the discretized ECG signals and corresponding saliency maps for each class was confirmed using Mutual Information (MI). We introduce the **Explainable-AI Trustworthiness (EAT)** framework, which specifies **operational and testable sufficiency criteria** for trustworthy ECG model explanations: (i) **informational grounding** to the expert ST–T segment, quantified by adjusted/normalized mutual information (**AMI/NMI**) with **time-aware permutation** nulls; (ii) **robustness** to **ST–T–restricted** counterfactual/adversarial edits (low flip-rate, small change in true-class probability Δptrue, and stable saliency); and (iii) **architectural faithfulness** (branch concordance or ablation consistency). We state and prove core results: **statistical soundness** (type-I error control and consistency of the composite test); **monotonicity** (increasing ST–T saliency mass cannot decrease AMI/NMI or Dice/IoU@k); an **equivalence** between ST–T reliance and lack of invariance to ST–T–bounded perturbations; and **branch faithfulness** for additive pre-fusion under a completeness condition (the total logit equals the sum of branch contributions). We then instantiate these tests on held-out ECG data. Under stated assumptions, EAT moves beyond subjective, post-hoc validation to **formal, falsifiable criteria** for clinical trustworthiness. This work signals a shift in ECG-AI interpretability: from agreement-only metrics to an **EAT-grounded** evaluation that couples information-theoretic dependence, architectural robustness, and clinically realistic perturbation tests, thereby influencing model selection and evaluation across datasets and providing a more reliable basis for clinical claims.

## 1. INTRODUCTION

Most neural network models in ECG signal processing rely on unimodal inputs [1], typically derived from limited features such as time-domain signals alone [2, 3, 4]. Some studies relied only on frequency domain features [5, 6, 7]. Other works were done using only time-frequency domain features [8, 9, 10].

The unimodal modeling strategy is prone to algorithmic bias, which limits generalizability in real-world clinical settings [11], where models disproportionately overfit a specific group of patient populations or clinical conditions. Such bias undermines diagnostic reliability and contributes to healthcare disparities [12], impeding the equitable deployment of AI in cardiology. Addressing this challenge requires a more holistic and domain-informed modeling strategy that accounts for the complementarity and synergies of information in diverse domains.

Building upon earlier work [13], which has leveraged mutual information to investigate redundancy and complementarity within multimodal ECG datasets for optimized model performance. This study uses the **Explainable-AI Trustworthiness (EAT) framework** to make explanation quality testable. We shift interpretability beyond visual saliency toward a **rigorous, data-driven assessment** with statistical guarantees, complemented by EAT's robustness and branch-faithfulness. The proposed model is designed to capture the necessary and sufficient features from the complete PQRST complex [14], thereby enhancing its capability to detect cardiovascular disease (CVD) with higher fidelity, reliability, and accuracy.

**Goal of Study**

The overarching goal of this study is to improve the reliability, interpretability, and fairness of ECG-based CVD diagnostics by designing a multimodal fusion architecture that unifies the synergistic strengths of the time, frequency, and time-frequency domains of the ECG signal. This approach is empirically validated to enhance clinical reliability and align with trustworthy AI principles in healthcare.

**Contribution**

Some of the significant contributions of the study to multiple domains include:

1. **Comprehensive Multimodal Model Benchmarking and Optimal Strategy Identification:** Twenty new multimodal and three unimodal deep learning models are developed and systematically evaluated for ECG-based classification of cardiovascular diseases. The intermediate fusion model (M4) emerged as the best model.

2. **Enhanced Robustness and Generalizability:** Performance consistency for robustness and generalizability was quantified based on Cohen's d effect sizes, and 95% confidence intervals for ΔF1 were estimated via bootstrap resampling.

3. **Improved Interpretability Fidelity Assessment:** Saliency maps were aligned with the raw ECG waveform, and the evaluation was based on divergence, sparsity, and attention variance. Standard sanity checks for saliency maps, including randomizing model weights and shuffling training labels, were also conducted. Saliency maps reliably reflected genuinely learned patterns rather than spurious input artifacts.

4. **Resilience to Noise:** Controlled ablation studies were done under noise conditions (e.g., Gaussian noise, baseline wander), demonstrating that the intermediate fusion model (M4) retained strong performance, further supporting its potential applicability in real-world clinical environments.

5. **EAT-Grounded, Quantifiable Explainability for Multimodal ECG:** We make AI explanation in cardiology quality testable rather than asserted by instantiating the EAT framework for ECG.

6. **Synthetic Data Physiological Plausibility Assessment:** The physiological plausibility methods include feature space analysis, visual inspection, evaluation of deep feature consistency, and Kullback-Leibler (KL) Divergence.

7. **Clinical Translation:** The experimental findings reveal that domain-based ECG multimodal models achieve high accuracy and impressive interpretability, while providing an enlightening perspective on the transformative potential of multimodal deep learning in cardiovascular disease diagnosis.

This paper is organized as follows: Section 1 discusses the introduction to the study. The methodology of the experiment is discussed in Section 2. Section 3 highlights the experiment. The discussion and conclusion of the study are presented in Sections 4 and 5, respectively.

## 1. METHODOLOGY

### 2.1. Overview

We systematically investigate both unimodal and multimodal deep learning approaches for the classification of cardiovascular diseases (CVDs) using electrocardiogram (ECG) signals. Time, frequency, and time-frequency domains were considered. The dataset used consists of four distinct classes: Normal, Myocardial Infarction (STEMI), History of STEMI, and Abnormal Heartbeat. We addressed class imbalance[15] using the Adaptive Synthetic Sampling (ADASYN) technique [16].

Each domain-specific representation was mapped to an architecture optimized for its structure and information characteristics. Time-domain signals were processed using a one-dimensional convolutional neural network (1D-CNN). For the time-frequency domain, continuous wavelet transform (CWT) was applied to convert ECG signals into spectrogram-like images, which were then analyzed using a two-dimensional convolutional neural network (2D-CNN) to extract spatial and temporal-spectral correlations. The frequency-domain signals, derived via Fast Fourier Transform (FFT), were input into a Transformer-based architecture to capture long-range dependencies and global frequency patterns.

### 2.2. Deep ECG Multimodal Fusion Pipeline

Figure 1 presents a generic ECG signal processing and multimodal fusion pipeline adopted in this study.

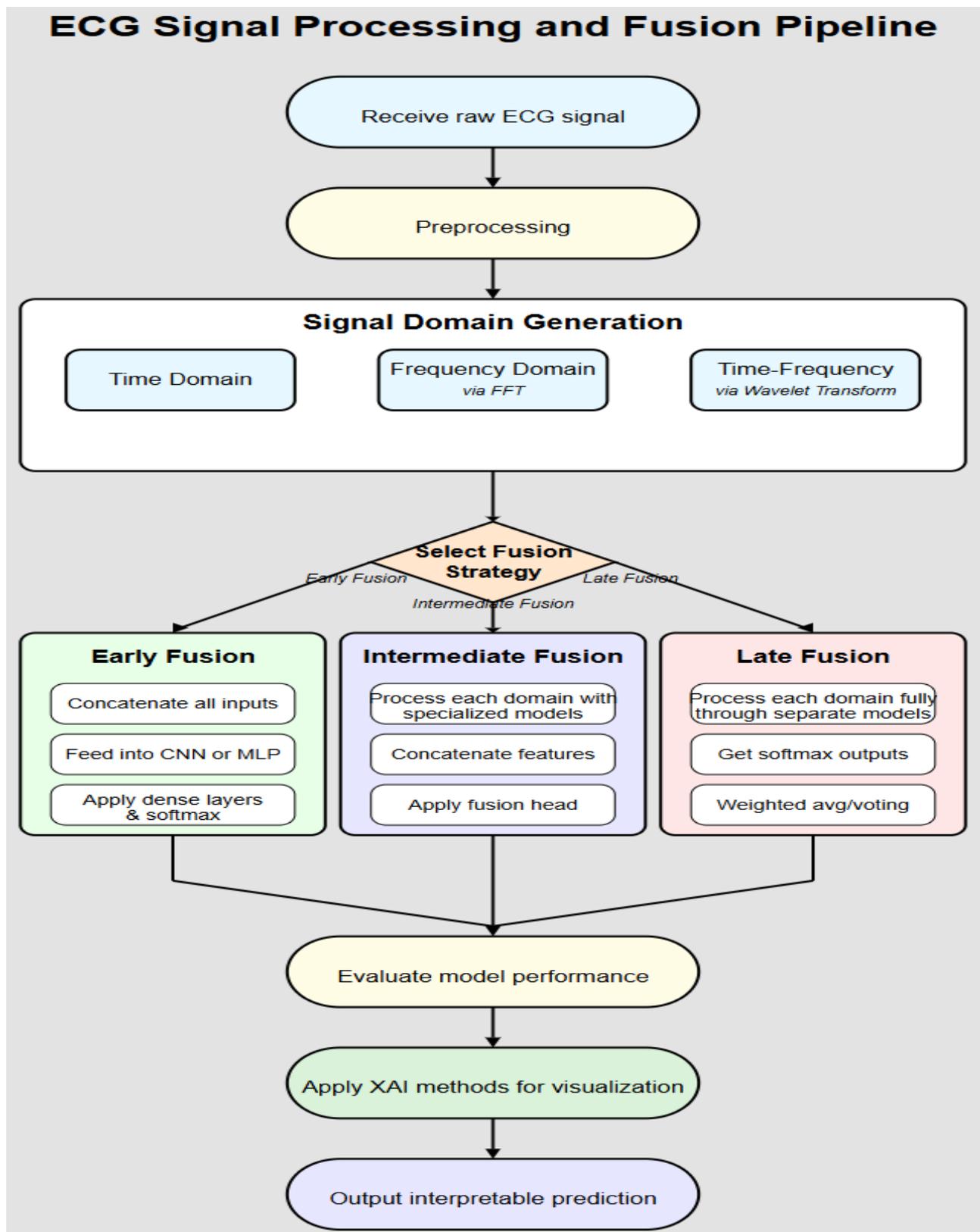

*Figure 1. ECG Signal Processing and Fusion Pipeline. Raw ECG signals are preprocessed to remove noise and filter signals outside a specified range.*

The workflow begins with the acquisition and preprocessing of raw ECG signals. Comparative performance evaluation is conducted across all fusion configurations to determine the most effective and clinically meaningful approach.

## 2.3. Dataset

The dataset used for this study was obtained from Mendeley Data [17].

**Summary of ECG Image Categories:**

- **Myocardial Infarction (STEMI):**
  240 patients × 12 images = 2,880 images
  → Exhibits ST-segment elevation and T-wave abnormalities.
- **Abnormal Heartbeat:**
  233 patients × 12 images = 2,796 images
  → Shows irregular heart rhythms like arrhythmias.
- **History of STEMI:**
  172 patients × 12 images = 2,064 images
  → Features altered QRS complexes and delayed conduction from past STEMI events.
- **Normal Individuals:**
  284 patients × 12 images = 3,408 images
  → Displays well-defined PQRST waveforms with minimal variations.

### 2.1. Handling Class Imbalance with Adaptive Synthetic Sampling (ADASYN)

Addressing class imbalance is crucial for building effective classification models, especially with medical datasets. In these datasets, minority classes often hold significant clinical importance. For instance, the ECG dataset in this study exhibited a clear imbalance, with the History of Myocardial Infarction class being significantly smaller than the other classes. To tackle this, the **Adaptive Synthetic (ADASYN) sampling approach** [18] was deployed to balance the training set. Choosing the correct algorithm to balance medical data is crucial due to its sensitive nature; an incorrect choice could have severe consequences.

ADASYN was selected over other oversampling techniques, such as traditional Synthetic Minority Over-sampling Technique (SMOTE) [19], due to its adaptive nature. It is particularly well-suited for the inherent complexities of ECG data. Unlike SMOTE, which generates synthetic samples uniformly along the line segments connecting minority class instances and their nearest neighbors, ADASYN focuses on generating more synthetic samples for minority class examples that are "harder" to learn [20]. It achieves this by adaptively shifting the decision boundary and generating more synthetic data points for samples located closer to the majority class or in sparse regions of the feature space. The sensitivity of medical datasets makes the adaptive density-based generation of ADASYN a suitable choice for this purpose.

ADASYN aims not just to balance the class distribution but also to improve the overall learning performance by focusing on regions that are challenging for the classifier, making it a highly suitable choice for the ECG dataset.

The effectiveness of ADASYN versus SMOTE on our dataset was empirically examined. Figure 2 shows that ADASYN outperformed SMOTE across all key metrics, including F1 score, accuracy, precision, and recall, while maintaining balanced performance across all classes. Given the clinical importance of consistent sensitivity in each class, ADASYN was selected as the more dependable oversampling strategy for this application.

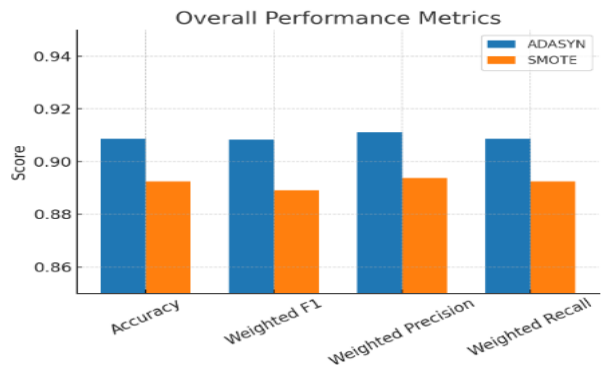

*Figure 2. Comparative performance of ADASYN and SMOTE.* **Comparative performance of ADASYN and SMOTE.**

Ultimately, ADASYN extends SMOTE by adapting the number of synthetic samples per instance based on the hardness of learning (i.e., local class imbalance) [21]. The idea is to focus more on ambiguous regions, improving the decision boundary. ADASYN was implemented using the following steps.

A. **Define local class imbalance around a minority sample $x_i$:**

$$r_i = \frac{\Delta_i}{k} \qquad (1)$$

Where:

- $\Delta_i$ is the number of $k$ nearest neighbors of $x_i$ that belong to the majority class
- $r_i \in [0,1]$ quantifies how "difficult" or "ambiguous" that point is

B. **Normalize difficulty weights:**

$$G_i = \frac{r_i}{\sum_{j=1}^{n} r_j} \cdot G \qquad (2)$$

Where:

o G is the total number of synthetic samples to generate
o $G_i$ is the number of synthetic samples to generate for point $x_i$

The mathematical formulation above is operationalized in Algorithm 1, which outlines the pseudocode for addressing class imbalance in the extracted time-frequency ECG representations. ADASYN adaptively generates synthetic examples in these regions to improve class balance and enhance model learning.

### Algorithm 1: ADASYN Balancing of Time-Frequency ECG Features

*I. Inputs:*
$\mathcal{D} = \{(I_i, y_i)\}$: Set of ECG scalogram images and class labels
$I_i \in \mathbb{R}^{\wedge}(H \times W)$: 2D ECG time-frequency image
$y_i \in \{1, 2, ..., C\}$: Class label

*II. Outputs:*
$X\_bal \in \mathbb{R}^{\wedge}\{N' \times s \times s\}$: Balanced scalogram matrix
$Y\_bal \in \{0,1\}^{\wedge}\{N' \times C\}$: One-hot encoded labels

*III. Parameters:*
*L: Total number of minority samples to generate*
*s: Desired image size after resizing (output resolution)*
*k: Number of nearest neighbors (default: 5)*

*IV. Step 1: Image-to-Signal Conversion*
*For each image $I_i$:*
$x_i(t) = (1 / W) \cdot \Sigma$ *(from w = 1 to W) $I_i(t, w)$*
*// Converts image row to 1D ECG signal by averaging columns*

*V. Step 2: Time-Frequency Representation*
*For each signal $x_i(t)$:*
$S_i(a, b) = \int x_i(t) \cdot \psi^*\_\{a, b\} (t) \, dt$ *// CWT transform*
$\mathcal{S}_i(a, b) = |S_i(a, b)|^2$ *// Power scalogram*
$\tilde{S}_i = Resize(\mathcal{S}_i, (s, s))$ *// Resize to fixed shape*

*VI. Step 3: Feature Matrix Construction*
$X = \{\tilde{S}_i\}, Y = \{y_i\}$ *for i = 1 to N*
$x_i = vec(\tilde{S}_i) \in \mathbb{R}^{\wedge}\{s^2\}$ *// Flatten scalogram*

*VII. Step 4: ADASYN Oversampling*
*For each minority class sample $x_i$:*
*Compute $d_i$: Degree of difficulty using k-NN*
$G_i = (d_i / \Sigma (j \in \mathcal{M}) \, d_j) \cdot G$ *// Adaptive weight*
*For each synthetic sample:*
*Choose neighbor $x_j$ from k nearest*
$\lambda \sim Uniform(0, 1)$
$x\_new = x_i + \lambda(x_j - x_i)$ *// Generate new point*

*VIII. Step 5: Reconstruction and One-Hot Encoding*
*Reshape $x\_new \in \mathbb{R}^{\wedge}\{s \times s\}$ to image format*
*Convert $y_i \in \{1, ..., C\} \rightarrow$ one-hot $\in \{0,1\}^{\wedge}C$*
*Return $X\_bal, Y\_bal$*

### 2.2. Validation of Synthetic Sample Physiological Plausibility

**Research question:** How can the physiological plausibility of the synthetic dataset generated by the ADASYN be quantitatively and qualitatively evaluated to ensure its suitability for downstream ECG modeling tasks?

To address this question, a multifaceted validation approach was employed to rigorously assess the quality of the ADASYN-generated synthetic scalograms. This approach included feature space analysis, visual inspection, evaluation of deep feature consistency, and Kullback-Leibler (KL) Divergence.

#### 2.2.1. Feature Space Distribution Analysis

The global distribution of both real and ADASYN-generated synthetic scalograms was assessed using t-distributed Stochastic Neighbor Embedding (t-SNE) [22, 23]. As shown in Figure 3, the synthetic samples consistently cluster within the high-density regions occupied by real samples across all four classes (Class 0–3). This spatial proximity in the reduced feature space indicates that the synthetic data preserves key statistical and structural properties of authentic ECG signals.

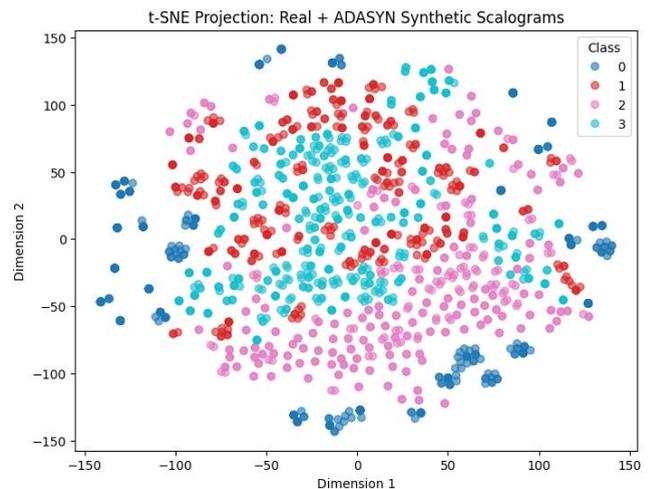

*Figure 3. Global Distribution of both Real and ADASYN-generated synthetic scalograms*

#### 2.2.2. Visual Fidelity to Physiological Patterns

To qualitatively assess signal realism, synthetic scalograms were visually compared to real ones [24]. Figure 4 shows side-by-side examples illustrating that synthetic samples replicate time-frequency characteristics, energy distributions, and structural patterns commonly observed in ECG signals. The resemblance supports the hypothesis that the ADASYN-generated scalograms retainphysiologically meaningful characteristics, while formal clinical validation remains outside the current scope.

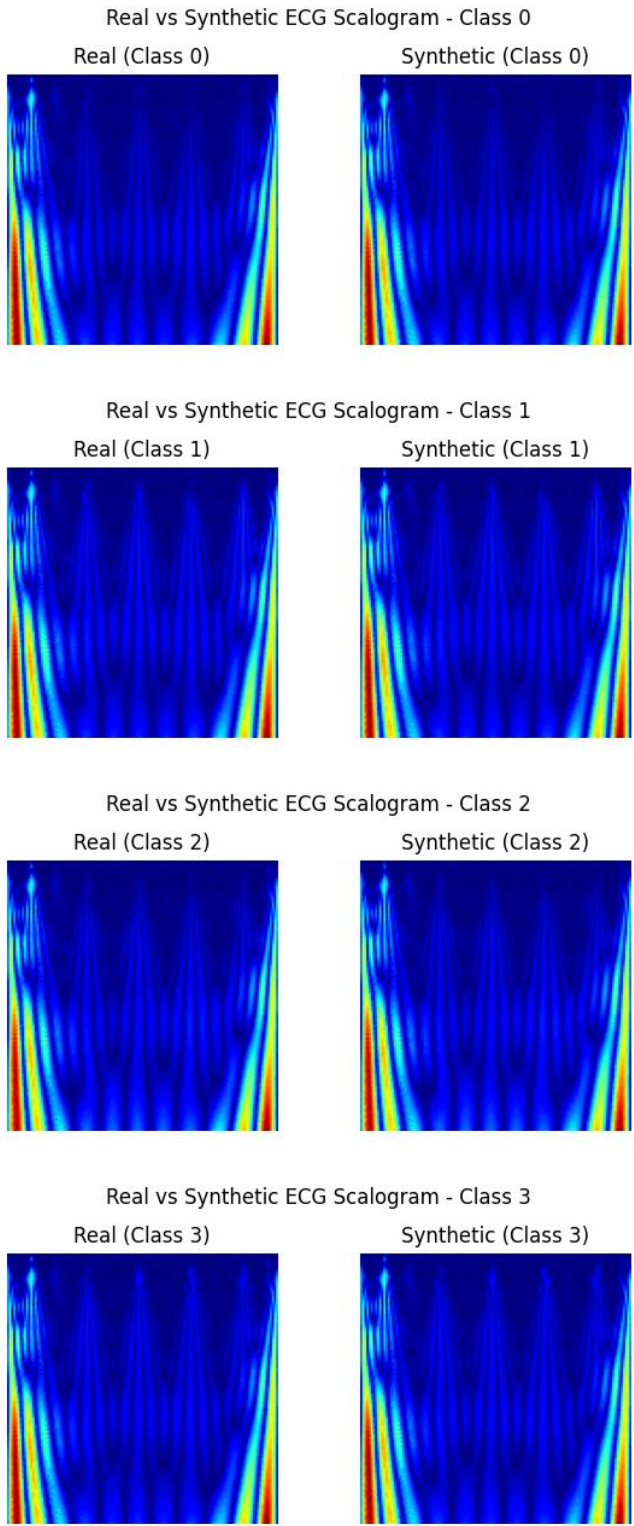

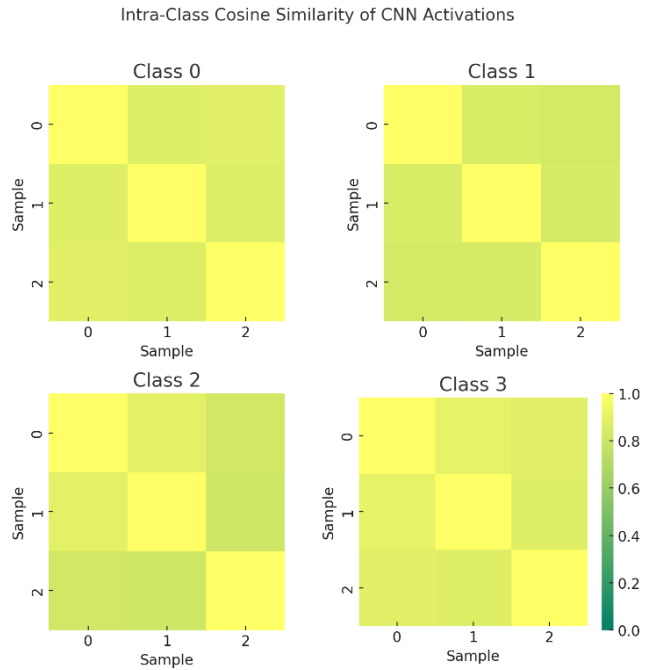

*Figure 5. Cosine Similarity Heatmap*

max_pooling2d_3 layer of the 2D-CNN. Intra-class cosine similarity was computed between real and synthetic samples, [25], with three examples per class. As visualized in Figure 5, real and synthetic samples elicited consistent deep activation patterns within each class. This substantial similarity in CNN-derived representations confirms that the synthetic data is encoded by the model in a manner consistent with real ECG inputs, supporting its utility for downstream training and generalization.

#### 2.2.4. KL Divergence

Finally, a statistical analysis using Kullback-Leibler (KL) Divergence was computed. A remarkably low KL Divergence of 0.0008 was observed. As KL Divergence quantifies the divergence of one probability distribution from another, a value nearing zero strongly indicates that the synthetic data's distribution is nearly identical to that of the real data [26, 27].

The converging qualitative and quantitative empirical evidence from the above complementary evaluations conclusively demonstrates that the generated synthetic dataset is physiologically plausible and highly suitable for downstream ECG modeling tasks.

### 2.3. Wavelet Decomposition

Wavelet decomposition is a signal processing technique that decomposes a signal into different frequency components and provides multi-resolution analysis [28].

Equation 3 represents the decomposition of signal S into its approximation (A) and detail (D) coefficients at different

*Figure 4. Visual comparison of real vs. synthetic ECG scalograms: morphology and energy distribution.*

#### 2.2.3. Consistency in Deep Feature Representation

To evaluate how a trained model interprets synthetic signals, feature representations were extracted from the levels (j). This process helps isolate important signal features and filter out noise.

$$S = A_j + \sum_{j=1}^{j} D_j \qquad (3)$$

Where:

- $A$ is the low-frequency approximation
- $D$ are the high-frequency detail coefficients at level j.

### 2.3.1. Thresholding for Noise Removal

Signals are decomposed to compute the detailed coefficients $D_j$ using soft thresholding.

$$D_j' = sign(D_j) . max(|D_j| - \lambda, 0) \qquad (4)$$

Where λ is the noise threshold.

As shown in equation 4, the choice of the threshold λ is crucial. Coefficients with an absolute value smaller than λ are set to zero, effectively removing noise. Coefficients with an absolute value greater than λ are shrunk, preserving the signal while reducing noise. Thus, the choice of λ is critical; a small λ may leave too much noise, while a large λ may remove important signal features [29].

### 2.3.2. Signal Reconstruction

After the detail coefficients have been thresholded, the inverse wavelet transform is applied to obtain the filtered time-domain signal [30]. Equation (5) shows the inverse wavelet transform, where the approximation coefficients at the final level and the modified detail coefficients are used to reconstruct the signal. This step effectively converts the processed wavelet representation back into a meaningful ECG signal, with the noise components significantly reduced.

$$S' = A_j + \sum_{j=1}^{j} D_j' \qquad (5)$$

### 2.3.3. Bandpass Filtering

Noises associated with the ECG dataset [31] was reduced using a Butterworth bandpass filter [32]. The Butterworth bandpass filter filtered out frequencies not within a specified range. Equation (6) represents the general bandpass filtering process.

$$H_{bandpass}(f) = H_{highpass}(f) - H_{lowpass}(f) \qquad (6)$$

The selected frequency range should encompass the dominant frequencies of the P-wave, QRS complex, and T-wave [33].

### 2.3.4. Ablation Study

**Research Question**: What is the optimal frequency range that balances effective noise suppression while preserving critical diagnostic information for deep learning-based ECG classification in this application?

We addressed the above question through a rigorous ablation study. Using a 1D Convolutional Neural Network (1D-CNN) on ECG signals preprocessed with various bandpass filter configurations, three specific bandpass filter settings was systematically evaluated. The goal was to pinpoint the range that maximizes classification accuracy and clinical utility.

A 1D-CNN architecture was chosen specifically for this analysis because of its proven ability to directly capture localized temporal features like QRS complexes, P-waves, and T-waves from raw or minimally processed ECG time-series data [34]. This architecture's efficiency, low computational complexity, and strong baseline performance were crucial for its selection, as they enabled controlled experiments where the sole impact of preprocessing steps could be isolated, free from architectural variability [35].

The three bandpass filtering configurations include:

I. A narrow 0.5–4.5 Hz filter focused on aggressive noise suppression (targeting baseline wander and high-frequency noise).

II. A standard 0.5–45 Hz filter commonly adopted in ECG signal processing for diagnostic tasks.

III. A wide 0.05–100 Hz filter enabling full-band signal admission, capturing both low-frequency drift and high-frequency detail.

The trained 1D-CNN model was evaluated across all three conditions. Confusion matrices and class-specific performance metrics revealed how filtering choices affect the model's ability to preserve clinically relevant information versus removing noise.

Figure 6 presents the corresponding confusion matrices for each configuration (ordered top to bottom: 0.5–4.5 Hz, 0.5–45 Hz, 0.05–100 Hz), providing empirical insight into the optimal trade-off between signal fidelity and denoising in ECG-based classification tasks.

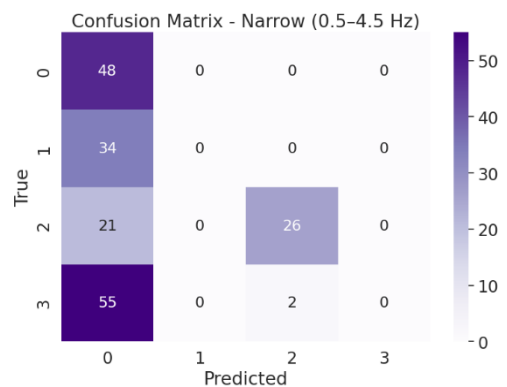

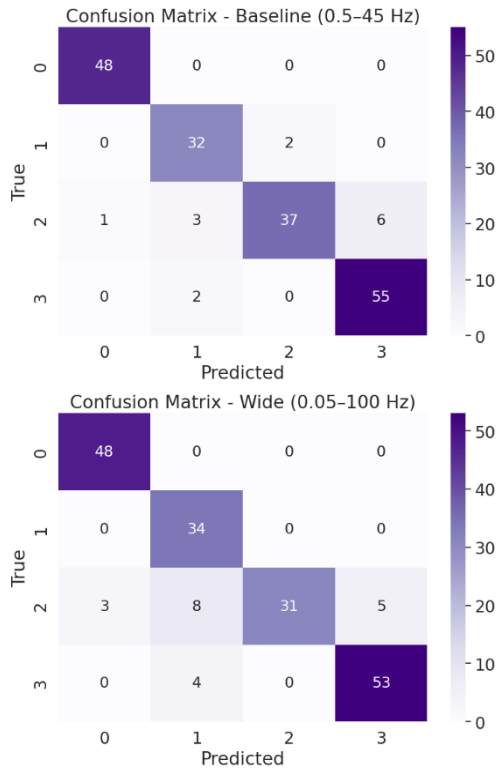

*Figure 6. Confusion matrices comparing ECG classification performance under three bandpass filter settings: (Top) Narrow filter (0.5–4.5 Hz), (Middle) Baseline filter (0.5–45 Hz), and (Bottom) Wide filter (0.05–100 Hz).*

Table 1 summarizes the interpretation of the experimental results.

*Table 1. Class-wise interpretation of filter performance and diagnostic trade-offs from the Confusion Matrix*

| Class | Narrow (0.5–4.5 Hz) | Baseline (0.5–45 Hz) | Wide (0.05–100 Hz) | Comment |
|---|---|---|---|---|
| **Class 0** | Accurate; QRS/ST-T well preserved | Accurate; QRS/ST-T well preserved | Accurate; QRS/ST-T well preserved | All filters provide clean separation and preserve morphology. |
| **Class 1** | Misclassified as Class 0; HRV features suppressed | Reduced precision | Correctly classified; RR and P variability retained | The narrow filter misclassified all. |
| **Class 2** | Heavily misclassified; ST/QRS slopes lost | High sensitivity; ST deviations | Reduced precision | ST elevation features are lost in the narrow |

| | | moderately retained | | filter; baseline performs best. |
|---|---|---|---|---|
| **Class 3** | All misclassified; morphology overly flattened | Consistently detected; voltage-based features preserved | Slightly degraded; minor distortion of voltage features | Wide and narrow filters degrade voltage-sensitive patterns; baseline stable. |

Performance evaluation based on Accuracy, Precision, Recall, and F1-Score is shown in Table 2. The baseline filter (0.5–45 Hz) demonstrated the best overall performance, striking a balance between noise suppression and signal preservation. The narrow band (0.5–4.5 Hz) led to substantial degradation across all metrics, likely due to excessive attenuation of clinically relevant ECG components (e.g., QRS complex and T wave). The wide range (0.05–100 Hz) preserved more information but admitted higher-frequency noise, slightly reducing generalization performance.

*Table 2. Performance Comparison Table*

| Filter Range | Accuracy | Precision | Recall | F1 |
|---|---|---|---|---|
| **Narrow (0.5Hz-4.5 Hz)** | 0.40 | 0.31 | 0.40 | 0.30 |
| **Baseline (0.5Hz-45 Hz)** | 0.92 | 0.93 | 0.92 | 0.92 |
| **Wide (0.05-100 Hz)** | 0.89 | 0.91 | 0.89 | 0.89 |

The results of the ablation study suggest a 0.5–45 Hz bandpass frequency as a pragmatic trade-off between artifact suppression and signal integrity, aligning with previous biomedical signal processing standards [36].

### 2.3.5. Butterworth vs. Adaptive Filters

**Research Question:** Does 0.5–45 Hz achieve the highest clean F1 and exhibit non-catastrophic behavior across common noise types compared with wide 0.05–100 Hz and adaptive filters?

Using identical test windows and the same 1D-CNN backbone, we compared a zero-phase Butterworth bandpass (0.5–45 Hz) with a wide Butterworth (0.05–100 Hz) and adaptive filters (ANC with pseudo-EMG reference; Kalman variants). Figure 7 shows our experimental results.

**Robustness analyses (Panels A–B)** favor 0.5–45 Hz: per-noise changes are small or nil (≈ −1% for Gaussian 15 dB, ~0% for baseline 0.15–0.3 Hz, ~0% for EMG 20–90 Hz), yielding a near-zero average ΔF1 across noises. Wide 0.05–100 Hz shows larger degradations (e.g., ~−7% for Gaussian and ~−5% for EMG on our dataset), while ANC is fragile (≈ −50% under Gaussian). Kalman variants sometimes show positive ΔF1 relative to their weak clean baselines, yet their **absolute** F1 under noise remains well below Butterworth's.

**Clean performance (Panel C)** further shows 0.5–45 Hz achieves the highest discriminability (F1_weighted ≈ 0.92; F1_macro ≈ 0.92), outperforming wide 0.05–100 Hz (≈ 0.88) and all adaptive alternatives (ANC ≈ 0.79; Kalman variants ≈ 0.54–0.55). **Confusion matrices (Panel D)** confirm a tighter diagonal and less cross-class leakage for 0.5–45 Hz.

The superior robustness and performance of the frequency range of **0.5–45 Hz** support the ablations (section 2.3.4) as the default preprocessing for the experiment.

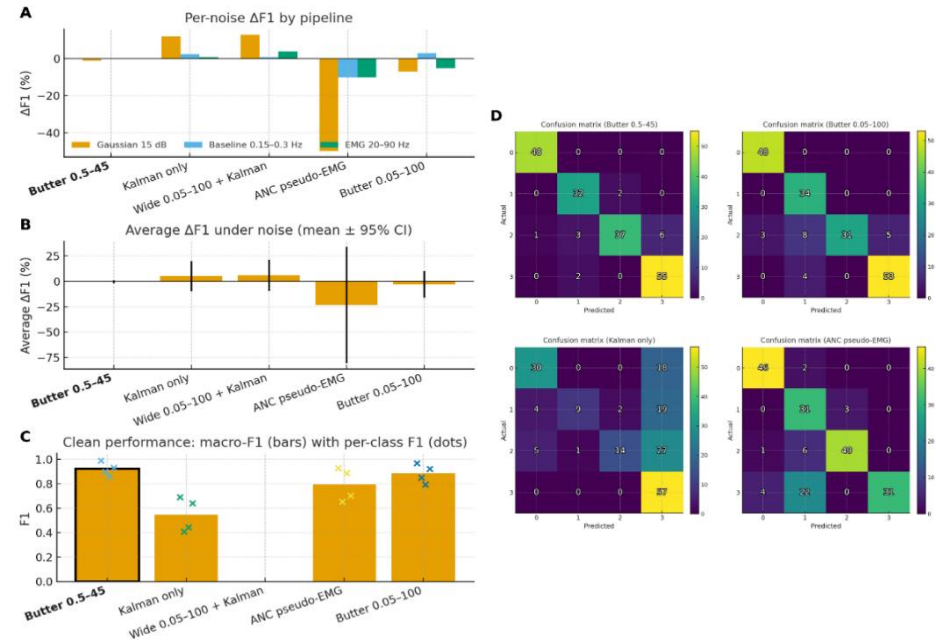

*Figure 7. Noise robustness across preprocessing pipelines*

### 2.3.6. QRS high-frequency preservation
**Research question.** Does a 0.5–45 Hz Butterworth filter effectively preserve diagnostically important QRS high-frequency content while suppressing noise?

Identical clean windows were filtered in two ways: zero-phase Butterworth (0.5–45 Hz) and wide (0.05–100 Hz). For each window, we calculated (i) QRS notch count (fQRS proxy), and (ii) band-limited RMS in 30–45 Hz (HF-ECG) and 45–90 Hz (very-HF/EMG overlap). Differences were defined as Δ = (Wide 0.05–100) − (Butter 0.5–45) and summarized as median [Q1–Q3]. For 30–45 Hz, we interpret Δ using a non-inferiority margin of 15% of the Butter median. Our result is shown in Figure 8.

Together with higher clean-set F1 and stronger noise robustness, our observations, as shown in Figure 8, support the idea that 0.5–45 Hz preserves the **HF morphology relied on by the model** (including fQRS patterns) while avoiding unnecessary very-HF/EMG energy.

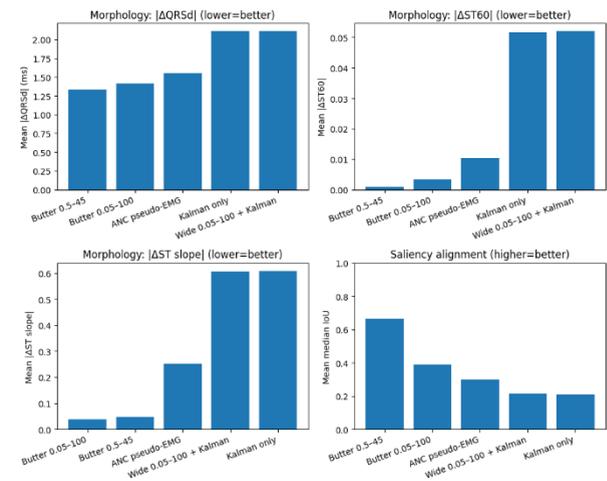

*Figure 8. QRS high-frequency (HF) preservation with a 0.5–45 Hz Butterworth front-end.*

Figure 9 illustrates the preservation of morphology and

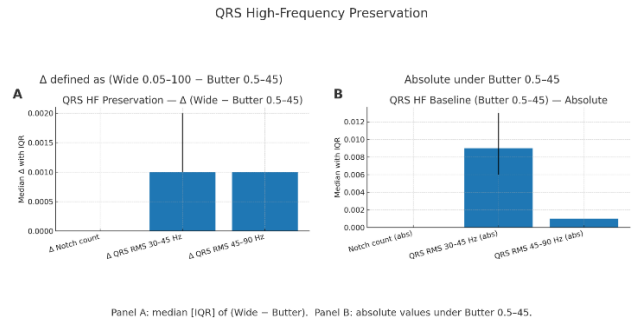

QRS High-Frequency Preservation

Panel A: median [IQR] of (Wide − Butter). Panel B: absolute values under Butter 0.5–45.

alignment of saliency across preprocessing pipelines.

*Figure 9. Morphology preservation and saliency alignment across preprocessing pipelines.*

Our experiments, including ablation results, robustness analyses, clean-data performance, and morphology preservation, collectively support **a filter range of 0.5–45 Hz as the optimal range for the study**.

### 2.4. Data Splitting Strategy

A stratified splitting strategy was employed using the StratifiedShuffleSplit function from the Scikit-learn library to split the data into training, testing, and validation sets. The dataset was first partitioned into an 80% training set and a 20% test set using stratified sampling to preserve the original class distribution. Subsequently, the 20% test set was further split evenly into two subsets, resulting in a 10% validation set and a 10% held-out test set. The stratification approach ensures that all subsets retain the same proportional distribution of class labels as the original dataset.

### 2.5. Signal Processing

Figure 10 illustrates the PQRST complex of an electrocardiogram (ECG) signal, representing the sequence of electrical events that occur during a single heartbeat [37]. This waveform is divided into distinct segments: the P wave, PR interval, QRS complex, ST segment, and T wave. Each corresponds to specific phases of cardiac depolarization and repolarization [38].

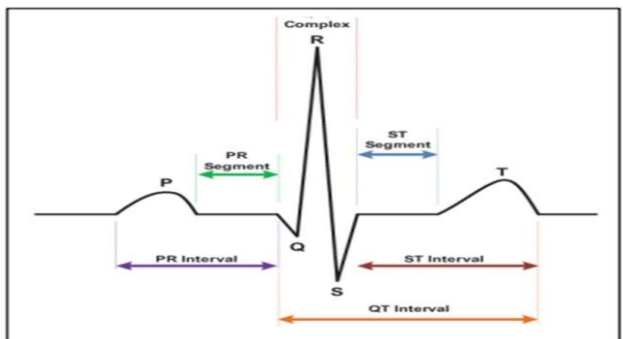

*Figure 10. The PQRST Complex [39]. The diagram depicts the P wave, QRS complex and T wave.*

In this study, signal processing techniques based on time, frequency, and time-frequency domains are applied, extracting feature representations from physiologically meaningful segments to support accurate classification of cardiovascular disease (CVD).

**Time Domain**

The QRS complex was evaluated in terms of width, amplitude, and electrical axis to identify potential ventricular conduction issues. Abnormalities in the ST segment, such as elevation or depression, suggest signs of myocardial ischemia or injury [40]. The T wave inversions or hyperacute peaks provided indicators of repolarization abnormalities [41]. To assess rhythm irregularities, RR intervals were computed for variability and arrhythmic patterns [42]. This comprehensive time-domain examination of raw ECG signals establishes a foundational basis for identifying a broad spectrum of cardiac pathologies, contributing to robust diagnostic decision-making [43].

#### 2.5.1. Frequency Domain

Fast Fourier Transform (FFT) and wavelet-based decomposition techniques were deployed [44]. The evaluation of R-R interval regularity in the frequency domain enhanced the sensitivity to detecting minor arrhythmic patterns, which may not be evident in the raw ECG signal. Spectral analysis of the ST segment and T wave provided complementary diagnostic insights, particularly for identifying myocardial ischemia and repolarization instability [45].

#### 2.5.2. Time-Frequency Domain

While time-domain analysis captures morphological variations in ECG signals and frequency-domain analysis reveals spectral energy distributions, both approaches individually fall short in preserving the dynamic temporal evolution of frequency content. To address this limitation, time–frequency domain feature extraction was employed, enabling simultaneous analysis of both time-localized and frequency-dependent signal characteristics. The ECG signal was segmented using a sliding window approach to track variations across key waveform components; the P wave, QRS complex, and T wave [46].

Each windowed segment underwent transformation using the Continuous Wavelet Transform (CWT) to generate scalogram images. It effectively illustrates how frequency content evolves over time [47]. This two-dimensional representation encodes both temporal localization and frequency scale resolution. It makes it suitable for capturing pathological variations in ECG morphology that are often subtle and nonstationary.

The resulting scalograms were treated as image-like inputs and processed using a two-dimensional Convolutional Neural Network (2D-CNN) to extract spatial patterns and texture information corresponding to time–frequency dynamics [48]. This approach facilitated the learning of both local and global features reflective of arrhythmic behavior, ischemic changes, or conduction abnormalities, especially in cases where traditional one-dimensional signal analysis might overlook such nuances [45].

### 2.6. Signal Visualization

To better understand the morphological and spectral characteristics of the ECG signals, Figure 11 presents representative visualizations across four classes, showing the transformation of the signal through successive stages: raw input, denoised and filtered output, time-domain representation, frequency-domain (FFT) spectrum, and time-frequency (wavelet scalogram) representation. The visualization of the ECG signals thus suggests that a single domain may be insufficient to fully capture the diagnostic variance in ECG signals, justifying the proposed multimodal fusion framework.

Table 3 presents a tabulated insight and knowledge derived from the visualized ECG signals across multiple domains. Each row represents a class, displaying a specific feature about the class. The feature type ranges from raw, time-domain data to complex time-frequency representations.

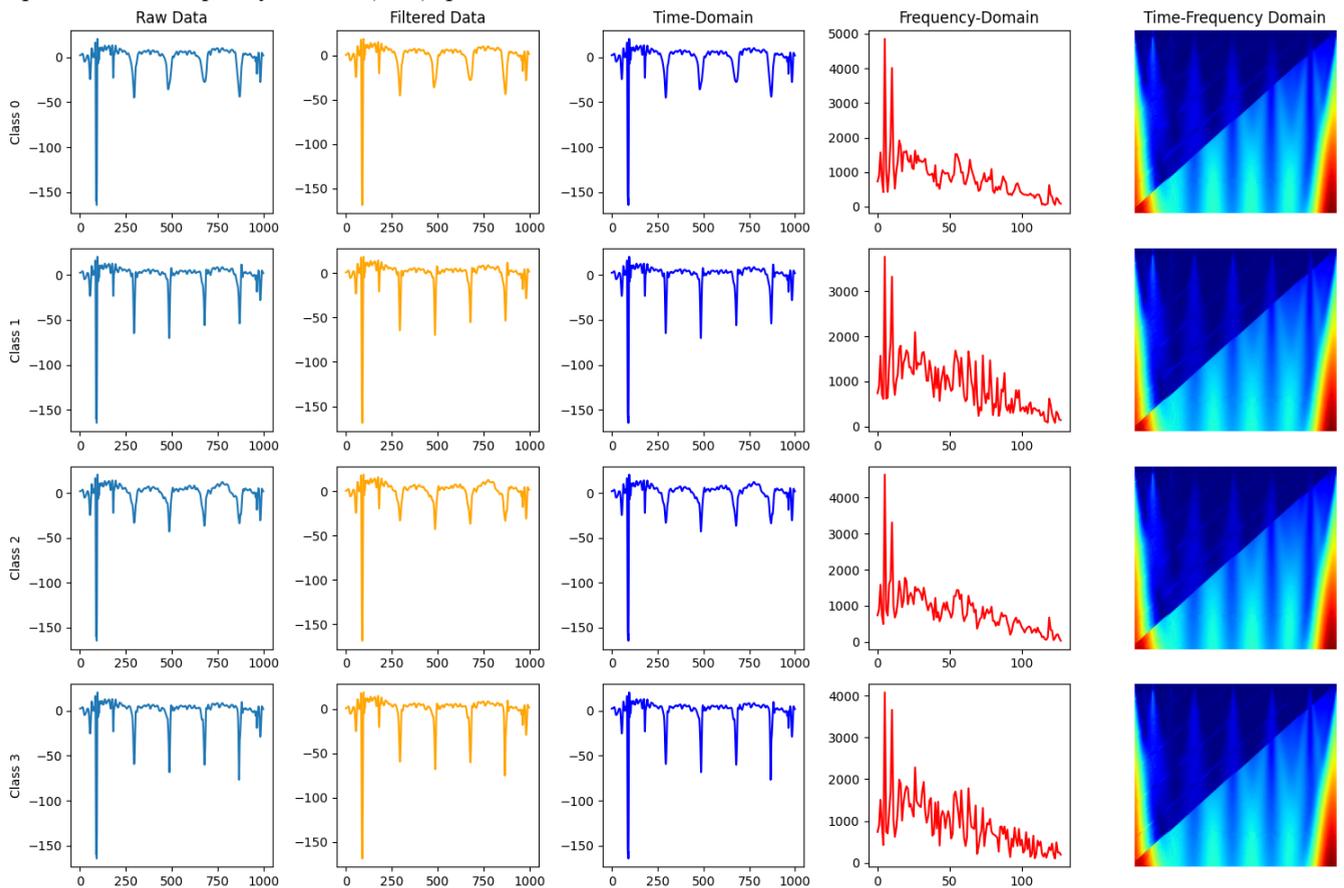

*Figure 11. Multistage ECG Signal Preprocessing and Multi-Domain Feature Representation Across CVD Classes. The figure illustrates the sequential transformation of ECG signals for four representative cardiovascular disease (CVD) classes (Class 0–3), progressing from raw to multi-domain formats used in deep learning.*

- **Column 1 (Raw Data):** Original ECG signals with baseline noise and artifacts.
- **Column 2 (Filtered Data):** Signals after Butterworth bandpass filtering to isolate clinically relevant frequency bands.
- **Column 3 (Time-Domain):** Clean waveforms capturing morphological features of PQRST complexes, used as input to 1D-CNN.
- **Column 4 (Frequency-Domain):** FFT-transformed signals highlighting dominant frequency components, fed to Transformer-based encoders.
- **Column 5 (Time-Frequency Domain):** Scalogram images derived via continuous wavelet transform, capturing both temporal and spectral features, used as input to 2D-CNNs.

*Table 3. Clinical Significance of ECG Feature Representation*

| Feature Type | Transformation Method | Clinical Relevance | Advantages | Limitations | Use in Diagnosis |
|---|---|---|---|---|---|
| **Raw Signal (Time-Domain)** | Direct ECG waveform recording | Captures electrical activity of the heart in real-time | Provides complete information, easy for manual interpretation | Contains noise (e.g., baseline wander, powerline interference) | Used by cardiologists for arrhythmia detection, myocardial infarction diagnosis [49] |
| **Filtered Signal (Time-Domain)** | Bandpass filtering (e.g., 0.5–45 Hz). | Reduces noise while preserving cardiac waveform morphology | Improves signal clarity, enhances feature extraction | May remove clinically relevant low-frequency components | Enhances ECG quality for automated anomaly detection [50] |
| **Time-Domain Features** | Feature extraction (e.g., PQRST intervals, HRV, RR interval). | Captures morphological characteristics of ECG beats | Effective for detecting arrhythmias, simple to compute | Limited ability to detect frequency-based abnormalities | Used for heart rate variability (HRV) analysis, arrhythmia classification [51] |
| **Frequency-Domain Features (FFT)** | Fast Fourier Transform (FFT) | Identifies dominant frequency components of ECG signals | Detects periodic heart abnormalities, robust to noise | Loses temporal information requires careful interpretation | Used in atrial fibrillation (AFib) detection, ischemic heart disease screening [52] |
| **Time-Frequency Features (Spectrograms)** | Wavelet Transform, STFT, CWT | Represents transient cardiac events in both time and frequency domains | Captures complex and transient patterns, good for deep learning models | Computationally intensive, requires high-quality data | Used for diagnosing ventricular tachycardia, ischemia, and transient arrhythmias [53] |

## 2.7. Fusion Strategies

**Research question:** Does the choice of fusion strategy (early, intermediate, or late) affect the classification performance of a multimodal deep learning architecture applied to an ECG dataset, considering the integration of time-domain, frequency-domain, and time-frequency domain features?

This study seeks to determine the optimal fusion strategy for multimodal ECG classification using deep learning. We systematically evaluate early, intermediate, and late fusion techniques.

### 2.7.1. Early Fusion

Early fusion involves the concatenation of features from multiple input domains at the initial stage of the model architecture prior to learning [54]. This approach ensures that the model learns from cross-domain interactions early in the learning pipeline. In the context of ECG-based cardiovascular disease (CVD) classification, early fusion typically merges features from:

- **Time-domain signals** (1D waveforms),
- **Frequency-domain components** (e.g., FFT-transformed vectors), and
- **Time-frequency representations** (e.g., wavelet-based scalograms or spectrograms).

Figure 12 illustrates an early fusion architecture for ECG signals, demonstrating how diverse domain features are merged at the input level and processed through a shared neural pipeline.

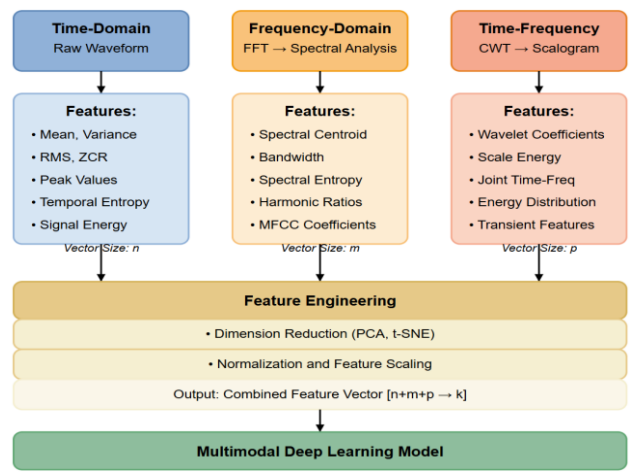

*Figure 12. Early Fusion Strategy.*

Early fusion serves as a valuable baseline for multimodal learning, offering insights into the combined feature space of ECG signals.

### 2.7.2. Intermediate Fusion

Unlike early fusion, which merges raw or preprocessed features at the input stage, intermediate fusion allows each modality to retain and refine its domain-specific characteristics through tailored processing pipelines prior to fusion [55]. Each domain is processed through a dedicated deep learning sub-network, such as:

- A **1D-CNN** for time-domain signal morphology,

- A **Transformer encoder** for frequency-domain spectral attention,
- A **2D-CNN** for spatial feature extraction from time-frequency images.

The significant disadvantage of this fusion strategy is its complexity, requiring more computational resources [56]. However, rigorous hyperparameter tuning and balanced training of the sub-models may neutralize some of these setbacks. Figure 13 illustrates the intermediate fusion pipeline, which involves independent domain-specific processing, followed by feature-level fusion and a final decision layer.

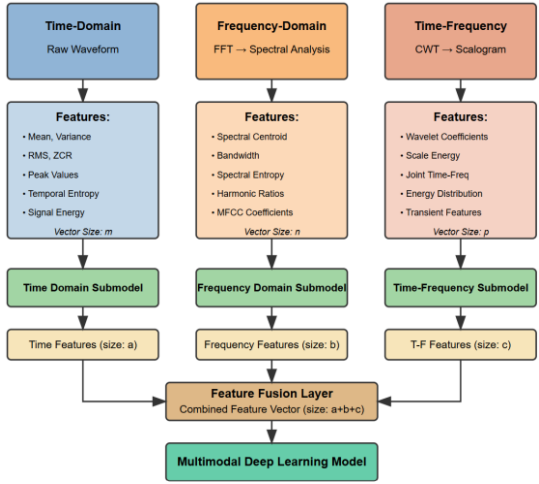

*Figure 13. Intermediate Fusion Strategy. The diagram demonstrates an intermediate fusion approach in multimodal deep learning.*

### 2.7.3. Late Fusion

Late fusion, also known as decision-level fusion, represents a modular and flexible approach where independent models are trained on each modality, and their predicted output (typically softmax probabilities) are later combined to make the final classification decision. Unlike early and intermediate fusion, late fusion bypasses the need to reshape or integrate raw or intermediate features during training. Instead, each model operates in isolation, optimizing its own representation of the data modality [57]. The final class prediction is derived by aggregating the output probabilities from each model using a weighted averaging strategy. For example, if all 3 models (1D-CNN, 2D-CNN and Transformer) are to be fused

$$y'_{fused} = \alpha_1 \cdot y'_{1D-CNN} + \alpha_2 \cdot y'_{2D-CNN} + \alpha_3 \cdot y'_{Transformer} \quad (7)$$

where:
- α1, α2, α3 are optimized fusion weights,
- $y'_{1D-CNN}, y'_{2D-CNN}$ and $y'_{Transformer}$ are the softmax probabilities from the individual models.

The weighted sum is followed by an argmax operation to determine the final predicted class label.

Figure 14 depicts the late fusion architecture, highlighting separate domain-specific branches and their final combination at the decision level.

**Key Characteristics of Late Fusion:**
- Simplicity in integration: Individual models can be developed, evaluated, and optimized independently.
- Modularity: Each branch can be swapped, retrained, or improved without retraining the entire system.
- Flexibility in domain contribution: Fusion weights can be adjusted based on validation accuracy or learned dynamically.

**Limitations:**
- The lack of joint feature learning may result in suboptimal performance compared to intermediate fusion, especially when cross-domain relationships are strong.
- Redundancy or dominance of a strong modality can skew the final decision if not carefully balanced.

Despite these limitations, late fusion has shown strong performance in various applications and provides a computationally efficient fallback when joint training is infeasible [58].

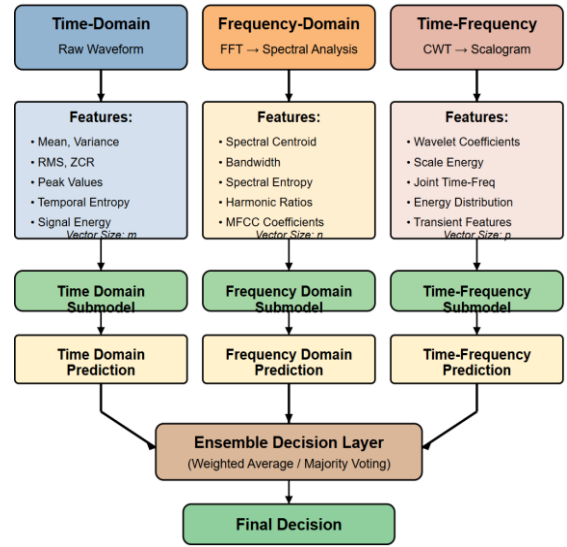

*Figure 14. Late Fusion Strategy. In this framework, features are extracted independently from the time, frequency, and time-frequency domains.*

## 3. Experiment

**Hypothesis**
*We hypothesize that multimodal data fusion across time, frequency, and time-frequency domains of ECG signals will significantly enhance predictive performance in cardiovascular disease (CVD) diagnosis with greater clinical interpretability.*

### 3.1. Unimodal Models

Three specific unimodal algorithms were designed:

- **Algorithm A**: 2D-CNN for the time-frequency domain
- **Algorithm B:** 1D-CNN for the time domain
- **Algorithm C**: 1D-CNN Transformer for the frequency domain.

A detailed exposition of Algorithms A, B, and C, including architecture and results, is presented in a related study [59]. Table 4 depicts the performance table.

*Table 4. Unimodal Performance Table*

| Model | Accuracy | Precision | Recall | F1-Score |
|---|---|---|---|---|
| **M1: 1D-CNN** | 0.94 | 0.94 | 0.94 | 0.94 |
| **M2: 2D-CNN** | 0.92 | 0.93 | 0.92 | 0.92 |
| **M3: Transformer** | 0.88 | 0.88 | 0.88 | 0.87 |

### 3.2. Early Fusion as a Baseline

Despite concerns about feature dilution caused by unaligned or redundant features, the performance of this approach was tested using a deep Multilayer Perceptron (MLP). The experimental setup for combining Time, Frequency, and Time-Frequency domains is shown in Figure 15. Unless otherwise noted, the hyperparameter values used are their default settings. s.

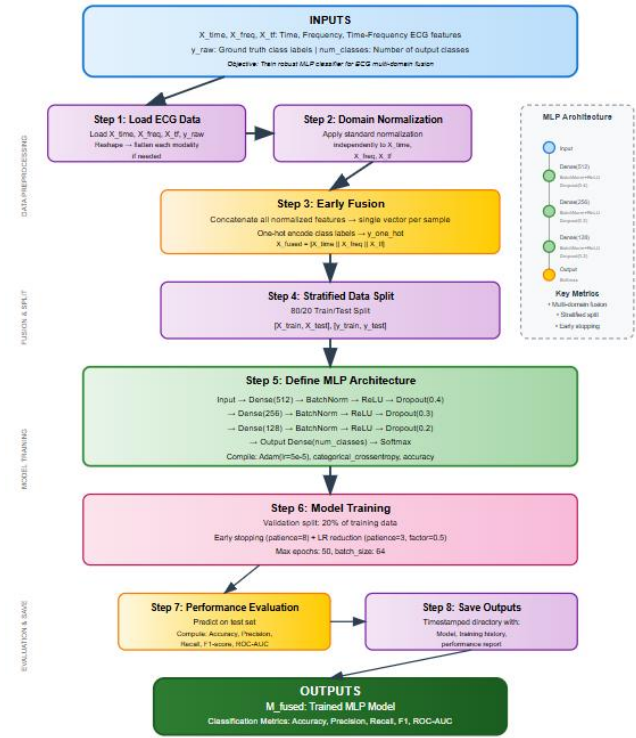

*Figure 15. Early Fusion-Based Multilayer Perceptron (MLP) for Multidomain ECG Classification Configuration. This framework*

integrates features extracted from three complementary domains: *time, frequency, and time-frequency into a unified feature representation.*

Table 5 summarizes the performance of different early fusion configurations evaluated using a deep MLP. The models fused features from two or three ECG domains (time, frequency, and time-frequency) and were evaluated using Accuracy, ROC-AUC, Macro F1-score, and Weighted F1-score.

As shown in Table 5, the early fusion MLP (Time + Time-Frequency) achieved an accuracy of up to 88.57% and a ROC-AUC of 0.9746 across several configurations. However, it was observed that adding more modalities (e.g., transitioning from two to three domains) did not significantly improve performance, suggesting saturation and potential feature redundancy. These findings reinforce the hypothesis that early fusion may suffer from feature dilution when combining heterogeneous representations, thereby motivating the design of more selective fusion mechanisms, such as intermediate and late fusion.

Consequently, early fusion serves as a crucial empirical baseline. Our experiment demonstrates the limitations of this naive feature aggregation and provides a contrast to more complex fusion strategies that explicitly model inter-modal interactions.

*Table 5. Performance Comparison of Early Fused Models*

| Model | Accuracy (%) | ROC-AUC | Macro Avg F1 | Weighted Avg F1 |
|---|---|---|---|---|
| Time + Time-Frequency | 0.8857 | 0.9746 | 0.88 | 0.88 |
| Time + Frequency | 0.8800 | 0.9635 | 0.88 | 0.88 |
| Time + Freq + Time-Freq | 0.8800 | 0.9645 | 0.88 | 0.88 |

Early fusion offers simplicity in implementation and a low-latency pipeline, but it introduces several limitations:

- Feature-level fusion prior to domain-specific abstraction may lead to dilution of domain-specific pattern**s** or interference among modalities.
- The approach lacks granular control over which domain contributes most to the prediction.
- The heterogeneity and misalignment of the different ECG signal domains (time, frequency and time-frequency) make fusion complex and potentially distorting information.
- Fusing domains together at the early stage can increase the dimensionality and complexity of the dataset. This raises concerns about computational cost and overfitting of the model.
- Early fusion is prone to noise which can impede the effectiveness of the model to learn necessary and

sufficient features for an efficient CVD predictive model.

Given the challenges of early fusion, this study focuses on intermediate and late fusion techniques. These two approaches are strategically chosen because they provide more robust mechanisms for leveraging domain-specific learning, especially in complex tasks like multimodal learning. By processing each modality through dedicated pipelines before integration (intermediate fusion) or by combining decisions from independently trained models (late fusion), these strategies improve the model's ability to capture and use complementary information from diverse data sources [62].

Figure 16 shows the experimental design of the study.

### 3.3. Intermediate Fusion Implementation
To implement intermediate fusion, learning algorithms were designed, trained and evaluated using fusion of features learned by the unimodal algorithms. By combining information after a learning algorithm has processed it, intermediate fusion exhibits greater robustness to noise [63]. The reduction of noise helps in mitigating overfitting, thereby improving the model's ability to generalize.

This experiment yielded four distinct intermediate fusion models:

- **M4:** Combining the learned features from the 1D-CNN (time domain) and the 1D-CNN Transformer (frequency domain). **(Algorithm B (features) + Algorithm C (features)).**
- **M5:** Combining the learned features from the 2D-CNN (time-frequency domain) and the 1D-CNN Transformer (frequency domain). (**Algorithm A (features) + Algorithm B (features)**)
- **M6:** Combining the learned features from the 1D-CNN (time domain) and the 2D-CNN (time-frequency domain). (**Algorithm A (features) + Algorithm C (features)**).
- **M7:** Combining the learned features from all three trained algorithms. (**Algorithm A (features) + Algorithm B (features) + Algorithm C (features)**)

#### 3.3.1. Post-hoc Fusion
The architectural details of our post-hoc intermediate fusion models are presented in Table 6. The experimental results are presented in both tabular and graphical forms in Table 7 and Figure 17, respectively. The results of the intermediate fusion show that M4, a combination of a 1D-CNN trained on the time domain and a Transformer trained on the frequency domain, achieves the best performance. The worst-performing model was the M2, a 2D-CNN trained on time-frequency domain features. Steps to implement multimodal MD4: 1D + Transformer are shown in Algorithm 2. This algorithm is strategically chosen because it has the best performance during the implementation of all intermediate models. Unless otherwise specified, hyperparameter values are their default values.

*Table 6. Intermediate Fusion Architecture*

| Model | Input Domains | Branch Components |
|---|---|---|
| **M4: 1D + Transformer** | Time (1D-CNN) Frequency (Transformer) | 1D-CNN: Conv1D → ReLU → Pool → BatchNorm → Flatten Transformer: FFT → Positional Encoding → MHA → FFN → Flatten |
| **M5: 2D + Transformer** | Time-Frequency (2D-CNN) Frequency (Transformer) | 2D-CNN: Conv2D → ReLU → Pool → Flatten Transformer: FFT → MHA → FFN → Flatten |
| **M6: 1D + 2D CNN** | Time (1D-CNN) Time-Frequency (2D-CNN) | 1D-CNN: Conv1D → ReLU → Pool → Flatten 2D-CNN: Conv2D → ReLU → Pool → Flatten |
| **M7: 1D + 2D + Transformer** | Time (1D-CNN) Time-Frequency (2D-CNN) Frequency (Transformer) | 1D-CNN: Conv1D → Pool → Flatten → Dense 2D-CNN: Conv2D → Pool → Flatten → Dense Transformer: FFT → Positional Encoding → MHA → FFN → Flatten |

# Multi-Domain ECG Classification Experimental Design

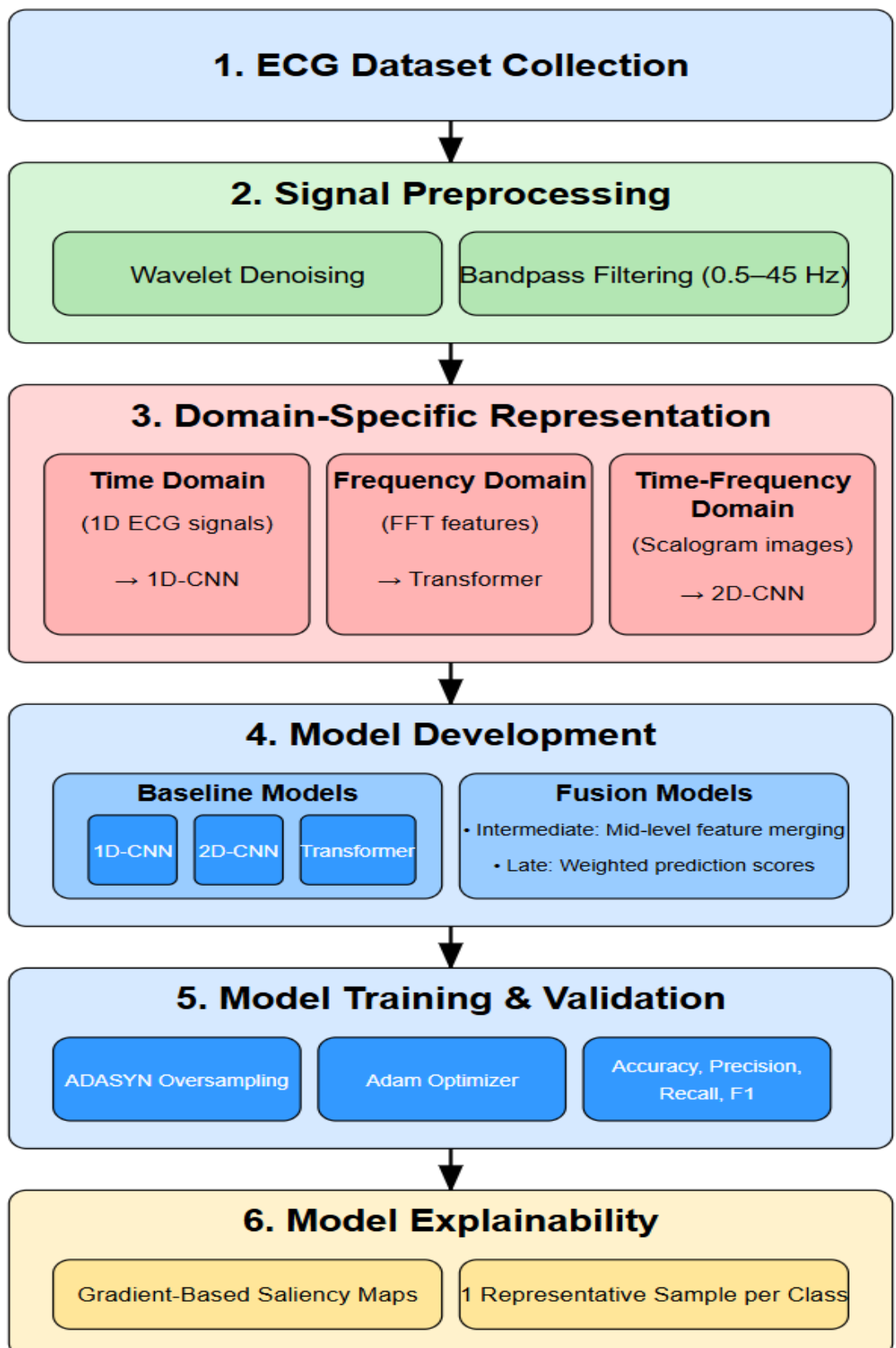

*Figure 16. Overview of the experimental design for the multi-domain ECG signal classification*

**Algorithm 2: Intermediate Fusion Model (MD4: 1D-CNN + Transformer)**

*Inputs:*
   *M_1D, M_Trans: Pretrained 1D-CNN and Transformer models*
   *X_1D, X_FFT: Time-domain and Frequency-domain data*
   *y_train, y_test: Ground truth class labels*
   *num_classes: Total number of output classes*

*Outputs:*
   *M_fused: Trained hybrid model*
   *Classification metrics: Accuracy, Precision, Recall, F1-score*
*Steps:*
*1. Load and preprocess training data:*
   *- Load X_1D, X_FFT, y_train*
   *- Reshape:*
      *X_1D → (N, 1000, 1)*
      *X_FFT → (N, 128, 1)*
   *- Align all arrays to equal length*
   *- Encode y_train to one-hot vectors → y_train_cat*

*2. Split into training and validation:*
   *- Perform stratified 80/20 split:*
      *[X_train_1D, X_val_1D], [X_train_FFT, X_val_FFT], [y_train_cat, y_val_cat]*

*3. Restore pretrained models:*
   *- Load M_1D and M_Trans*
   *- Set M_1D.trainable ← True*
   *- Set M_Trans.trainable ← True*

*4. Build intermediate fusion model:*
   *- Define input_1D ← Input(shape=(1000, 1), name="1D_ECG")*
   *- Pass through M_1D → Flatten → Dense(256, ReLU) → BatchNormalization → x1*

   *- Define input_FFT ← Input(shape=(128, 1), name="Transformer_ECG")*
   *- Pass through M_Trans → Flatten → Dense(256, ReLU) → BatchNormalization → x2*

   *- Concatenate(x1, x2) → merged*
   *- Dense(512, ReLU) → BatchNormalization → Dropout(0.25) → output*

   *- Build M_fused ← Model([input_1D, input_FFT], output)*
   *- Compile M_fused with:*
      *Optimizer ← Adam(learning_rate=3e-5)*
      *Loss ← categorical_crossentropy*
      *Metric ← accuracy*

*5. Train M_fused:*
   *- Use early stopping on val_loss with patience=10*

   *- Fit model on:*
      *Inputs: [X_train_1D, X_train_FFT]*
      *Labels: y_train_cat*
      *Validation: ([X_val_1D, X_val_FFT], y_val_cat)*
      *Epochs: 60*
      *Batch size: 32*

*6. Evaluate performance:*
   *- Load and preprocess X_test_1D and X_test_FFT*
   *- Predict: y_pred ← argmax(M_fused.predict([X_test_1D, X_test_FFT]))*
   *- Compute metrics:*
      *Accuracy, Precision (weighted), Recall (weighted), F1-score (weighted)*

*7. Return:*
   *Trained model M_fused and classification report*

*EndAlgorithm*

*Table 7. Performance Comparison Table of Intermediate Fusion Models*

| Model | Accuracy | Precision | Recall | F1-Score |
|---|---|---|---|---|
| **M4: 1D + Transformer** | **0.97** | **0.97** | **0.97** | **0.97** |
| M5: 2D + Transformer | 0.89 | 0.90 | 0.89 | 0.89 |
| M6: 1D + 2D CNN | 0.95 | 0.95 | 0.95 | 0.95 |
| M7: 1D + 2D + Transformer | 0.95 | 0.95 | 0.95 | 0.95 |

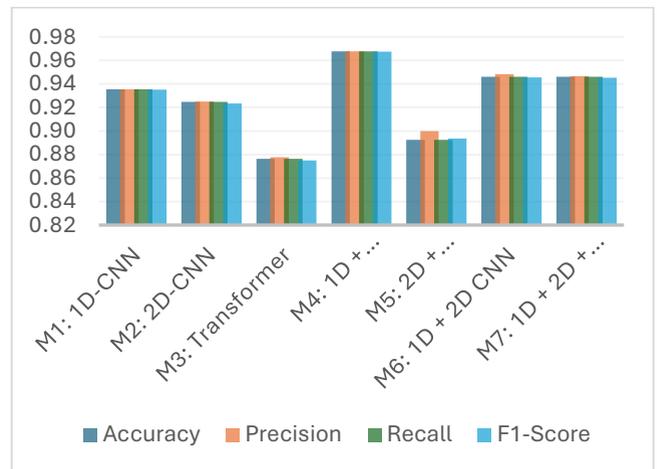

*Figure 17. Performance comparison of all evaluated models (M1–M7) across four classification metrics: accuracy, precision, recall, and F1-score.*

### 3.3.2. Cross-Attention
To model inter-domain interactions beyond post-hoc feature concatenation, we introduced bidirectional cross-attention blocks at the feature level. Tables 8 and 9 illustrate the model's architecture and comparison results of the models, respectively.

*Table 8. Intermediate Cross-Attention Fusion Architecture*

| Model | Input Domains | Branch Components | Fusion |
|-------|--------------|-------------------|--------|
| **M4-XA (1D + TX)** | Time (1D-CNN), Frequency (Transform er) | 1D: Conv1D → ReLU → MaxPool (tokens). TX: FFT → Dense proj → PosEnc → MHA → FFN. | Bidirectional cross-attention (1D↔TX; 4 heads; d=128) → GAP per stream → Concat → Dense → Softmax |
| **M5-XA (2D + TX)** | Time–Freq (2D-CNN), Frequency (Transform er) | 2D: Conv2D → ReLU → MaxPool tokenize (H×W→tokens). TX as above. | Bidirectional cross-attention (2D↔TX) → GAP → Concat → Dense |
| **M6-XA (1D + 2D)** | Time (1D-CNN), Time–Freq (2D-CNN) | 1D: Conv1D → ReLU → MaxPool (tokens). 2D as above. | Bidirectional cross-attention (1D↔2D) → GAP → Concat → Dense |
| **M7-XA (1D + 2D + TX)** | Time, Time–Freq, Frequency | 1D & 2D as above; TX as above. | Hierarchical: Stage-1 (1D↔TX) → pooled seed; Stage-2 (seed↔2D) → Concat → Dense |

*Table 9. Performance Comparison Table of Intermediate Cross-Attention Fusion Models*

| Model | Accuracy | Precision | Recall | F1-Score |
|-------|----------|-----------|--------|----------|
| **M4: 1D + Transformer** | 0.89 | 0.88 | 0.88 | 0.88 |
| **M5: 2D + Transformer** | 0.81 | 0.80 | 0.80 | 0.79 |
| **M6: 1D + 2D CNN** | 0.90 | 0.90 | 0.90 | 0.89 |
| **M7: 1D + 2D + Transformer** | 0.90 | 0.89 | 0.89 | 0.89 |

The performance comparison of the best models of the two different intermediate configurations is shown in Table 10.

Table 10. Best Intermediate Fusion Strategy Model Comparison

| Method (Family) | Top Config (Modaliti es) | Best Test Acc | Best Prec | Best Recall | Best F1 | Mean F1 |
|-----------------|--------------------------|---------------|-----------|-------------|---------|---------|
| Post-hoc | M4: 1D + Transfor mer | 0.97 | 0.97 | 0.97 | 0.97 | 0.94 |
| Cross-Attention Fusion | M6: 1D + 2D CNN | 0.90 | 0.90 | 0.90 | 0.89 | 0.86 |

As shown in Table 10, the Post-hoc concatenation of the intermediate is the better configuration. Across M4–M7, **cross-attention did not yield a measurable improvement over simple feature concatenation** under parameter-matched training (full results in Table 9). In contrast, **modal choice** was decisive: models that included both 1D (temporal morphology) and 2D (time–frequency texture) consistently outperformed configurations that relied solely on the frequency Transformer, rather than a 1D branch.

### 3.3.3. Late Fusion Implementation
Late fusion was implemented by aggregating the prediction probabilities of independently trained unimodal models using a weighted averaging scheme [64]. The goal was to determine sets of fusion weights that maximize the overall classification performance of the combined model.

#### 3.3.3.1. Grid Search Technique
An exhaustive search approach using the grid search technique was employed by systematically exploring combinations of weights across the prediction vectors. This method is widely used in multimodal fusion literature due to its simplicity and effectiveness in identifying near-optimal fusion parameters without requiring gradient-based optimization [65, 66]. The resulting fused model benefited proportionally from the complementary strengths of the individual modalities, enhancing robustness and diagnostic accuracy in ECG signal classification [67] [3].

The specific steps for late fusion in this study were:
- Train individual base models on their respective input domains:
  - A 2D Convolutional Neural Network (2D-CNN) trained on scalogram image (Algorithm A)
  - A 1D Convolutional Neural Network (1D-CNN) trained on time-domain data (Algorithm B)
  - A Transformer network trained on Fast Fourier Transform (FFT) signals (Algorithm C)

- For each test sample, the softmax output (probability distribution over classes) from each trained model was saved:
  - 1D-CNN saved as y_pred_1d.npy.
  - 2D-CNN saved as y_pred_2d.npy.
  - Transformer saved as y_pred_transformer.npy,

- The final prediction for each sample was obtained by calculating a weighted sum of these saved softmax outputs.
- A grid search technique was used to systematically explore different combinations of weights ($\alpha$) to identify the set of weights that resulted in the best overall performance of the fused model.
- This optimization process was conducted exclusively on the independent validation dataset to

prevent data leakage and mitigate the risk of overfitting the fusion weights. Once the optimal weights were determined based on their performance on the validation set, these fixed weights were then applied to the completely unseen test dataset for the final performance evaluation of the late fusion models (e.g., M7).

- The rigorous separation ensures that the reported performance metrics for late fusion models accurately reflect their generalization capabilities on new, unobserved data.

Architectural details of the Late Fusion models are shown in Table 11. Figures 18 to 21 show the grid search graph for the optimal fusion of the models. Experimental results are presented in Table 12 and Figure 22, in both tabular and graphical representations. The results indicate that M7: 1D-CNN + 2D-CNN + Transformer achieves the best performance. M5: 1D-CNN + Transformer and M6: 2D-CNN + Transformer are tied for the worst performance. Algorithm 3 outlines the logical steps for implementing M7. This algorithm was strategically chosen because it has the best performance during implementation. Unless otherwise specified in the algorithm, hyperparameter values are their default values.

*Table 11. Late Fusion Architecture*

| Model | Modalities Used | Architecture Description |
|---|---|---|
| **M4: 1D-CNN + 2D-CNN** | Time + Time-Frequency | Classification probabilities from each model are aggregated: $\hat{y} = \alpha_1\hat{y}1D + \alpha_2\hat{y}2D$ $\alpha_1 = 0.7, \alpha_2 = 0.3$ |
| **M5: 1D-CNN + Transformer** | Time + Frequency | Independent predictions from 1D-CNN and Transformer. Final decision based on weighted averaging: $\hat{y} = \alpha_1\hat{y}1D + \alpha_2\hat{y}Transformer$ $\alpha_1 = 0.6, \alpha_2 = 0.4$ |
| **M6: 2D-CNN + Transformer** | Time-Frequency + Frequency | Softmax outputs from 2D-CNN and Transformer are combined: $\hat{y} = \alpha_1\hat{y}2D + \alpha_2\hat{y}Transformer$ $\alpha_1 = 0.7, \alpha_2 = 0.3$ |
| **M7: 1D-CNN + 2D-CNN + Transformer** | Time + Time-Frequency + Frequency | All three models predict independently. Final prediction is derived from: $\hat{y} = \alpha_1\hat{y}1D + \alpha_2\hat{y}2D + \alpha_3\hat{y}Transformer$ $\alpha_1 = 0.3, \alpha_2 = 0.3, \alpha_3 = 0.4$ |

Notably, to avoid overfitting during fusion weight optimization, the grid search was conducted exclusively on a validation subset derived from the training data. The final performance metrics, including accuracy, precision, recall, and F1-score, were evaluated on an independent test set held out throughout training and fusion tuning. This separation ensures that the reported improvements reflect the model's generalization performance rather than validation bias.

**Algorithm 3: Multimodal Weighted Late Fusion (M7: 1D-CNN + 2D-CNN + Transformer)**

*Inputs:*
    *M_1D, M_2D, M_Trans: Trained 1D-CNN, 2D-CNN, and Transformer    models (From Algorithms A, B, and C)*
    *X_1D, X_2D, X_FFT: Time, Time-Frequency, and Frequency-domain test data*
    *y_true: Ground truth class labels*
    $\alpha_i \in [0.0, 1.0]$ *such that* $\alpha_1 + \alpha_2 + \alpha_3 = 1$

*Outputs:*
    *Optimal weights ($\alpha_1^*$, $\alpha_2^*$, $\alpha_3^*$)*
    *Best classification metrics: Accuracy, Precision, Recall, F1-score*

*Steps:*
*1. Load trained models:*
    *M1D ← M_1D*
    *M2D ← M_2D*
    *MTrans ← M_Trans*

*2. Load and align test inputs:*
    *Load X_1D, X_2D, X_FFT, y_true*
    *Ensure all inputs have equal sample length*

*3. Predict class probabilities from each model:*
    *Ŷ_1D ← M1D(X_1D)*
    *Ŷ_2D ← M2D(X_2D)*
    *Ŷ_Trans ← MTrans(X_FFT)*

*4. Initialize empty list: results ← []*

*5. For $\alpha_1$ in [0.0, 0.05, …, 1.0]:*
*6.      For $\alpha_2$ in [0.0, 0.05, …, 1.0]:*
*7.           $\alpha_3$ ← 1.0 - $\alpha_1$ - $\alpha_2$*
*8.           If $\alpha_3 \in [0.0, 1.0]$:*
*9.                Ŷ_fused ← $\alpha_1$·Ŷ_1D + $\alpha_2$·Ŷ_2D + $\alpha_3$·Ŷ_Trans*
*10.               y_pred ← argmax(Ŷ_fused, axis=1)*
*11.               Compute metrics: Accuracy, Precision, Recall, F1-score*
*12.               Append ($\alpha_1$, $\alpha_2$, $\alpha_3$, Accuracy, Precision, Recall, F1 score) to results*

*13. Select the tuple with the best Accuracy:*
    *($\alpha_1^*$, $\alpha_2^*$, $\alpha_3^*$, Acc*, Prec*, Rec*, F1*) ← argmax(results)*

*14. Return optimal weights and evaluation metrics*

*EndAlgorithm*

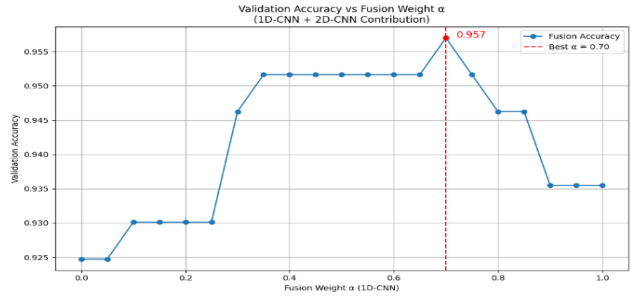

Figure 18. Figure 18. M4: 1D-CNN + 2D-CNN. This result highlights the synergistic value of both domains while affirming the dominance of time-domain signals in achieving higher classification accuracy.

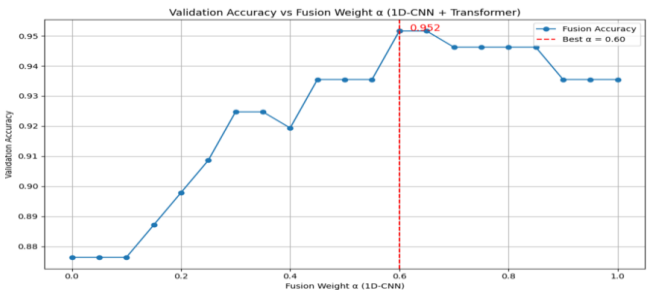

Figure 19. M5: 1D-CNN + Transformer. The fusion weight α ∈ [0, 1] controls the relative contribution of the 1D-CNN prediction in the final decision: α = 0 uses only the Transformer output, while α = 1 uses only the 1D-CNN. The plot shows that the optimal accuracy (0.952) is achieved at α = 0.60, as indicated by the red vertical dashed line.

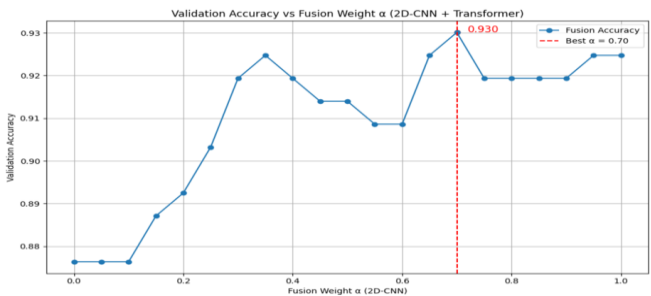

Figure 20. M6: 2D-CNN + Transformer. The fusion weight α ∈ [0, 1] controls the proportion of the final prediction influenced by the 2D-CNN, with (1 − α) contributed by the Transformer. The highest validation accuracy of 0.930 occurs at α = 0.70, as indicated by the red dashed line.

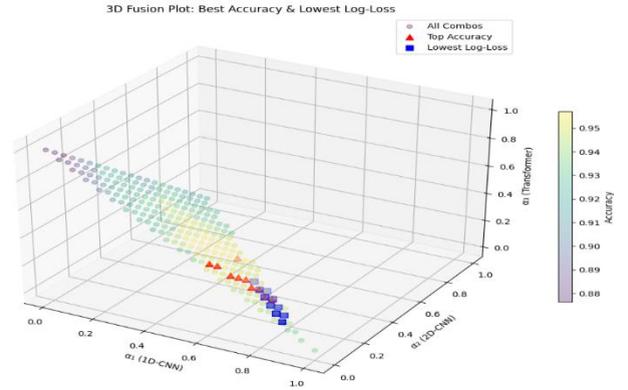

Figure 21. M7 :1D-CNN + 2D-CNN + Transformer. 3D Visualization of fusion weight combinations ($α_1$: 1D-CNN, $α_2$: 2D-CNN, $α_3$: Transformer) and their corresponding classification accuracies. The optimal validation accuracy of 0.96 is attained when the $α_1$ values of 1D-CNN (Time domain), 2D-CNN (Time-Frequency domain), and Transformer (Frequency domain) are 0.3, 0.3, and 0.4, respectively. with the lowest log-loss, reflecting better model confidence.

Table 12. Performance Comparison Table of Grid Search Late Fusion Models

| Model | Accuracy | Precision | Recall | F1-Score |
|---|---|---|---|---|
| 1D + 2D | 0.94 | 0.94 | 0.94 | 0.94 |
| 1D + Transformer | 0.93 | 0.94 | 0.93 | 0.93 |
| 2D + Transformer | 0.93 | 0.93 | 0.93 | 0.93 |
| 1D + 2D + Transformer | **0.96** | **0.96** | **0.96** | **0.96** |

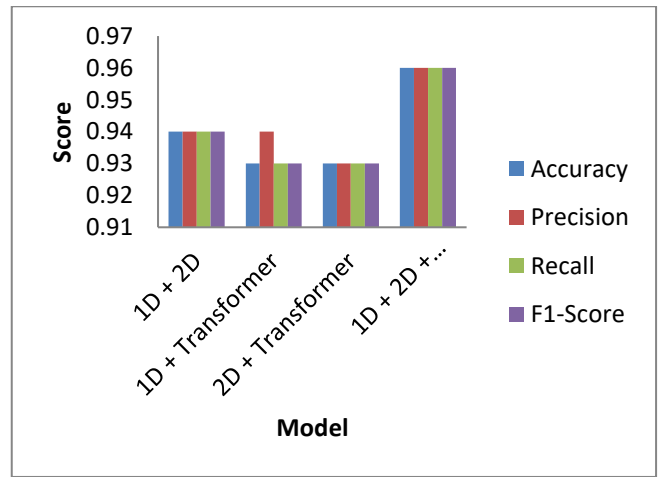

Figure 22. Comparison of performance metrics accuracy, precision, recall, and F1-score—across four hybrid ECG classification models.

To gain further insight into the optimization of late fusion for this study, we explored the Learned Attention Head (per-sample weights) and Learned Class-wise Weighting with Entropy Gating. The architectures of the two strategies are shown in Tables 13 and 15, respectively. Results of their implementation are shown in Tables 14 and 16.

Table 13. Learned class-wise weighting x entropy gating

| Model | Branches (Encoder -> Domain) | Fusion per class c (before renorm) |
|-------|------------------------------|-------------------------------------|
| M4 | 1D-CNN -> Time; 2D-CNN -> Time-Freq | w_1D,c * g_1D(x) * p_1D(c\|x) + w_2D,c * g_2D(x) * p_2D(c\|x) |
| M5 | 1D-CNN -> Time; Transformer -> Freq | w_1D,c * g_1D(x) * p_1D(c\|x) + w_TX,c * g_TX(x) * p_TX(c\|x) |
| M6 | 2D-CNN -> Time-Freq; Transformer -> Freq | w_2D,c * g_2D(x) * p_2D(c\|x) + w_TX,c * g_TX(x) * p_TX(c\|x) |
| M7 | 1D-CNN -> Time; 2D-CNN -> Time-Freq; Transformer -> Freq | w_1D,c * g_1D(x) * p_1D(c\|x) + w_2D,c * g_2D(x) * p_2D(c\|x) + w_TX,c * g_TX(x) * p_TX(c\|x) |

*Table 14. Performance Comparison of Learned Class-wise Weighting + Entropy Gating Models*

| Model | Accuracy | Precision | Recall | F1-Score |
|-------|----------|-----------|--------|----------|
| 1D + 2D | 0.95 | 0.94 | 0.94 | 0.94 |
| 1D + Transformer | 0.94 | 0.93 | 0.93 | 0.93 |
| 2D + Transformer | 0.93 | 0.92 | 0.92 | 0.92 |
| 1D + 2D + Transformer | 0.94 | 0.93 | 0.94 | 0.94 |

Table 15. Learned attention head (per-sample weights) Architecture

| Model | Modalities Used | Architecture Description |
|-------|-----------------|--------------------------|
| M4: 1D-CNN + 2D-CNN | Time + Time-Frequency | Two-branch attention fuser. Inputs are the calibrated class-probability vectors from 1D-CNN and 2D-CNN. Fusion head: concatenate probabilities → Dense(64, ReLU) → Dropout(0.2) → Dense(2) → Softmax to produce per-sample modality weights that sum to 1. Final class probability is the weighted sum of the two branch probabilities. The fusion head is trained with cross-entropy; backbones are frozen. |
| M5: 1D-CNN + Transformer | Time + Frequency | Same attention fuser as M4, but with 1D-CNN and Transformer probabilities. The head learns per-sample weights (two outputs with Softmax) and fuses via weighted sum of branch probabilities. Trained with cross-entropy; backbones remain frozen. |
| M6: 2D-CNN + Transformer | Time-Frequency + Frequency | Two-branch attention fuser over 2D-CNN and Transformer probabilities. Concatenate → Dense(64, ReLU) → Dropout(0.2) → Dense(2) → Softmax to obtain per-sample modality weights; weighted sum yields final probabilities. Only the fusion head is trained. |
| M7: 1D-CNN + 2D-CNN + Transformer | Time + Time-Frequency + Frequency | Tri-modal attention fuser. Inputs are three calibrated probability vectors. Fusion head: concatenate → Dense(64, ReLU) → Dropout(0.2) → Dense(3) → Softmax to produce per-sample weights for all modalities; weighted sum forms the final probabilities. Backbones are frozen; only the fusion head is trained. |

Table 16. Learned attention head (per-sample weights) Architecture Models

| Model | Accuracy | Precision | Recall | F1-Score |
|-------|----------|-----------|--------|----------|
| M4: 1D + 2D | 0.94 | 0.94 | 0.94 | 0.94 |
| M5: 1D + Transformer | 0.92 | 0.92 | 0.92 | 0.92 |
| M6: 2D + Transformer | 0.94 | 0.93 | 0.93 | 0.93 |
| M7: 1D + 2D + Transformer | 0.93 | 0.92 | 0.92 | 0.92 |

Table 17 compares the three late-fusion strategies we evaluated.

Table 17. Late Fusion Configuration Comparison

| Method (Family) | Top Config (Modalities) | Best Test Acc | Best Prec | Best Recall | Best F1 | Mean F1 |
|---|---|---|---|---|---|---|
| **Grid Search (fixed convex weights)** | **M7: 1D + 2D + Transformer** | **0.96** | **0.96** | **0.96** | **0.96** | **0.94** |
| **Learned class-wise weighting + entropy gating** | **1D + 2D** | 0.95 | 0.94 | 0.94 | 0.94 | 0.93 |
| **Learned attention head (per-sample weights)** | **M4 or M6 (tie)** | 0.94 | 0.94 | 0.94 | 0.94 | 0.93 |

Overall, as shown in Table 17, **M7 with grid-searched fixed weights** offers the **best result.** Therefore, we adopt it as the **preferred late-fusion configuration**.

## 4. DISCUSSION

Figure 23 summarizes the framework used to structure the multimodal ECG classification study. The workflow was organized into three sequential stages: (i) modality extraction, (ii) fusion strategy design, and (iii) derivation of key findings. To identify the optimal fusion approach, a comparative analysis was conducted between the highest-performing models from each strategy.

Figure 24 highlights the architectural comparison between the best intermediate fusion model (M4: 1D-CNN + Transformer) and the best late fusion model (M7: 1D-CNN + 2D-CNN + Transformer).

In M4, the latent feature representations extracted from the 1D-CNN (which captures temporal features) and the Transformer (which captures frequency-based dependencies through self-attention) were concatenated to form a unified representation. This fused feature vector was then fed into a fully connected classification head, allowing the model to learn joint representations and optimize decision boundaries using combined temporal-spectral information.

In contrast, M7 adopts a late fusion strategy. Each modality is processed independently:

- 1D-CNN learns from raw temporal sequences,
- 2D-CNN processes time-frequency scalograms,
- Transformer extracts frequency-domain representations.

Each branch produces an independent softmax prediction. Predictions were then aggregated using a weighted summation, where the contribution of each modality is controlled by learned or manually tuned weights. While this design allows for modularity and model specialization, it does not exploit joint feature interactions across modalities, which may limit its capacity to capture higher-order cross-domain dependencies.

This study deliberately adopted the weighted fusion approach for M7, as it enables selective emphasis on more informative modalities, rather than using uniform averaging which may suppress dominant signal cues.

The performance comparison presented in Figure 24 supports the hypothesis that intermediate fusion (as implemented in M4) more effectively captures complementary features and leads to improved classification performance over late fusion (M7), particularly in the context of complex ECG signal variability across CVD classes.

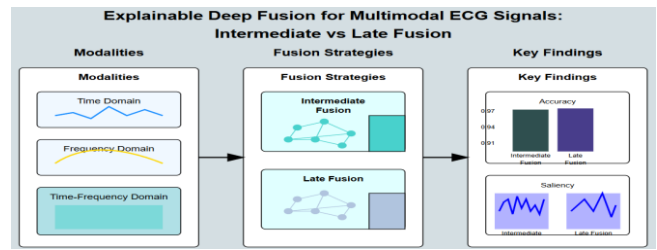

*Figure 23. Architectural model comparison diagram illustrating the framework and findings of the experiment.*

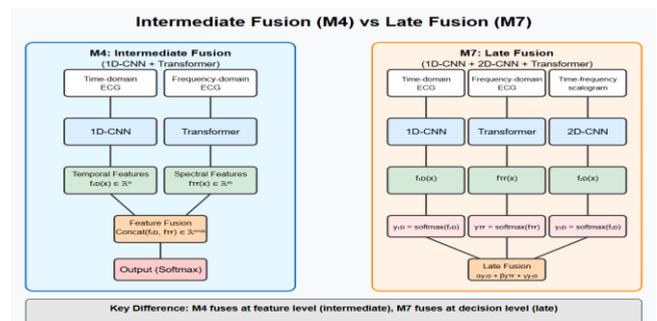

*Figure 24. Best Intermediate Fusion (M4) VS Best Late Fusion (M7).*

### 4.1. Training Configuration and Reproducibility

All models (unimodal and multimodal) in this study were implemented in Python using the Keras deep learning library with a TensorFlow backend. Unless otherwise noted, training was performed using the Adam optimizer ($\beta_1 = 0.9$, $\beta_2 = 0.999$, $\varepsilon = 1e{-}7$) with an initial learning rate of 0.001 and a batch size of 64. A learning rate schedule was applied using Keras's LearningRateScheduler callback to reduce the rate by a factor of 0.1 every 10 epochs. Training ran for a maximum of 100 epochs, with early stopping based on validation loss (patience = 10) and automatic restoration of the best weights. The source code will be released on GitHub upon manuscript acceptance to support transparency and reproducibility.

### 4.2. Theoretical Premises (Mathematical Statements)

Let:

$X_{TD}$: Time domain features (1DCNN)
$Z_{TD}$ : Latent Space for IDCNN
$f_{ID}$: $X_{TD} \rightarrow Z_{TD}$
$X_{FD}$: Frequency domain features (Transformer)
$Z_{FD}$ : Latent Space for Transformer
$F_{FD}$: $X_{FD} \rightarrow Z_{FD}$
$X_{TFD}$: Time-Frequency domain features (2DCNN)
$Z_{TFD}$: Latent Space for 2DCNN
$F_{TFD}$: $X_{TFD} \rightarrow Z_{TFD}$
$Ý_1$ : Prediction from the 1DCNN
$Ý_2$ : Prediction from the 2DCNN
$Ý_3$ : Prediction from the Transformer
$Ý_{LT}$: Final Prediction from the Late Fusion
$Ý_{IT}$: Final Prediction from the Intermediate Fusion
$Y$: Target
$g_{IT}$ : Intermediate Fusion Classifier
$I(A,B)$: Mutual information between $A$ and $B$
$h_{IT}$ : Intermediate Fusion Transformation
$f$ : a combined function

#### 4.2.1. M4: 1DCNN + Transformer (Best Intermediate)

- *Fusion Layer:*
  $h_{IT}$: $Z_{TD} \times Z_{FD} \rightarrow Z_{IT\_}{}^{fused}$

- *Classification:*
  *The fused representation is passed through a classifier*

  $g_{IT}$:$Z_{IT\_}{}^{fused} \rightarrow Ý_{IT}$

- *Information Theoretic Perspective:*
  $I (Y ; Ý_{IT}) \leq I (Y; Z_{IT\_}{}^{fused})$

- *If joint information of $Z_{1D}$ and $Z_{FD}$ are required to predict Y, then,*
  - $I (Y; Z_{1D}, Z_{FD}) > I (Y; Z_{1D})$
  - $I (Y; Z_{1D}, Z_{FD}) > I (Y; Z_{FD})$

- *For an optimized intermediate fusion $Z_{IT}{}^{fused}$*
  - $I (Y; g_{IT} (h_{IT} (Z_{1D}, Z_{FD}))) = I (Y; Z_{1D} , Z_{FD})$

- *In practice, due to model capacity and training limitations, the achieved mutual information will be:*
  - $I (Y; Z_{IT\_fused}) \approx I (Y; Z_{1D}, Z_{FD}) - \epsilon$
  - *where $\epsilon \geq 0$ represents the information loss during the fusion and classification process*

#### 4.2.2. M7: 1D + 2D + Transformer (Best Late)

- *Independent modeling:*
  *Each subnet produces an independent prediction:*
  $Z_{TD} \rightarrow Ý_1$
  $Z_{TFD} \rightarrow Ý_2$
  $Z_{TF} \rightarrow Ý_3$

- *Fusion at Decision Level:*
  $Ý_{LT} = \sum_i w_i . Ý_i$

- *Information Theoretic Interpretation:*
  $I (Y; Ý_{IT}) \leq f (I (Y; Ý_1), I (Y; Ý_2), I (Y; Ý_3))$

Figure 25 presents the comparative performance of the best intermediate and late fusion strategies using Accuracy, Precision, Recall, and F1-score as evaluation metrics. The results clearly show that the best intermediate fusion model, M4 (1D-CNN + Transformer), consistently outperforms the best late fusion model, M7 (1D-CNN + 2D-CNN + Transformer). This superior performance of M4 may be attributed to its fusion strategy, which operates at the feature level. By jointly learning from temporal (1D-CNN) and spectral (Transformer) latent representations prior to classification, M4 can capture richer, more complementary information. The concatenation of these latent embeddings enables the model to form a more expressive joint representation, which enhances class separability and reduces ambiguity during classification.

In contrast, M7 applies fusion at the decision level, where each modality-specific model (1D, 2D, Transformer) makes an independent prediction, and the final decision is derived through weighted aggregation. most likely, M7 might not have benefited from synergetic relationship among the models. Therefore, the advantage of M4 stems from its ability to perform deep integration of modalities during feature learning, resulting in better generalization and improved clinical prediction accuracy.

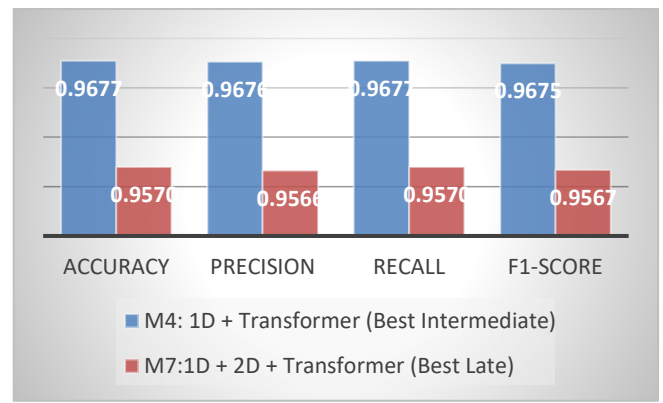

*Figure 25. Comparison of classification performance between the best intermediate fusion model (M4: 1D-CNN + Transformer) and*

the best late fusion model (M7: 1D + 2D + Transformer). Comparison across accuracy, precision, recall, and F1-score metrics.

### 4.3. Noise Robustness Evaluation for M4: 1DCNN + Transformer (Best Intermediate)

**Research Question**: Can the intermediate fusion model (M4: 1D-CNN + Transformer) maintain satisfactory performance (F1-score) when subjected to independent injections of Gaussian noise (15 dB SNR), baseline wander (0.15 Hz), and muscle noise (20–50 Hz), thereby demonstrating its robustness for practical ECG classification?

To evaluate the robustness of the intermediate fusion model (M4: 1D-CNN + Transformer) against real-world ECG signal disturbances, a controlled ablation study was performed by introducing three types of clinically relevant noise into the test dataset. These selected noise types represent common artifacts seen in ambulatory ECG recordings.

- Gaussian noise to simulate environmental interference (15 dB SNR) [68].
- Baseline wander to emulate motion-induced signal drift (0.15 Hz) [69].
- Muscle noise to capture high-frequency electrical activity from muscle contractions (20–50 Hz) [70].

Each noise condition was tested separately, and model performance was re-evaluated using the F1-score to measure degradation compared to clean test data. This experiment mimics real-world deployment scenarios, where model robustness against signal corruption is essential for clinical reliability. The steps to carry out noise evaluation in the proposed model (M4) are detailed in Algorithm 4.

**Algorithm 4: Noise Robustness Evaluation for M4**

*Inputs:*
  *$M_4$: Pretrained intermediate fusion model (1D-CNN + Transformer)*
  *X_test: Clean ECG test samples*
  *y_test: Ground truth class5 labels*
  *fs: Sampling frequency*

*Outputs:*
  *F1_clean: Baseline F1-score under clean condition*
  *Δ: Degradation from clean baseline across noise types*

*Steps:*
1. *Define ECG noise types:*
    *noise_types ← {Gaussian, BaselineWander, Muscle}*

2. *Predict on clean test data:*
    *ŷ_clean ← $M_4$(X_test)*
    *F1_clean ← F1_score(y_test, ŷ_clean)*

3. *Initialize results ← []*

4. *For noise_type in noise_types:*
5.    *X_noisy ← ApplyNoise(X_test, noise_type, fs)*
6.    *ŷ_noise ← $M_4$(X_noisy)*
7.    *F1_noise ← F1_score(y_test, ŷ_noise)*
8.    *Δ ← F1_noise - F1_clean*
9.    *Append (noise_type, F1_noise, Δ) to results*

10. *Return:*
    *F1_clean, results*

*EndAlgorithm*

The results of implementing the above algorithm are shown in Table 18.

Table 18. Robustness of M4 to Noise Injection

| Noise Type | F1-score | Δ from Clean (%) |
|---|---|---|
| Clean (baseline) | 0.963 | — |
| Gaussian Noise (15 dB) | 0.942 | -2.18% |
| Baseline Wander | 0.934 | -3.01% |
| Muscle Noise | 0.926 | -3.84% |

In Table 18, we observed that the M4 model's performance remains good under all noise conditions. The F1-score reductions are all under 4%. This confirms the model is suitable for real-life applications for ECG monitoring, where noise pollution is a common challenge. Out of all the artifacts, muscle noise is the most dominant, causing a performance drop of 3.84%. This suggests that future work can investigate techniques such as data augmentation or adaptive filtering to further enhance model robustness.

### 4.4. Cohen'd Comparison of Intermediate and Late Fusion Models vs. Standalone Baselines for ECG Classification

**Research Question:** To what extent does intermediate fusion (M4) significantly outperform unimodal and late fusion models (M7) in ECG-based cardiovascular disease classification, as measured by Cohen's d, and performance metrics such as F1-score, accuracy, and precision?

The superior performance of fused models, both intermediate and late fusion, is consistently highlighted across all evaluation metrics, outperforming their unimodal counterparts. The empirical results support the hypothesis that multimodal fusion boosts the performance of deep learning models compared to models trained on single domains. To statistically confirm these performance improvements further, Cohen's d was calculated, revealing that the differences between fused and unfused models are statistically significant. Additionally, a comparison was made between the top-performing late fusion model and the best intermediate fusion model to determine which one is empirically dominant. The results indicate that multimodal fusion, especially when performed through intermediate or late integration, significantly enhances classification accuracy, robustness, and clinical reliability in ECG signal interpretation. [71, 72].

Effect size analysis using Cohen's *d* provides a robust and interpretable statistical measure for evaluating the practical significance of observed differences in model performance. While conventional metrics such as accuracy and F1-score indicate predictive effectiveness, Cohen's *d* quantifies the magnitude of performance improvement between fusion-based models and their unimodal counterparts, independent of sample size [73]. This distinction is fundamental in clinical machine learning, where statistically significant improvements must also demonstrate meaningful effect sizes to justify their application in sensitive diagnostic environments [74]

Each fusion model was compared with standalone models built on individual modalities (1D-CNN for the time domain, 2D-CNN for the time-frequency domain, and Transformer for frequency domain features) in the context of cardiovascular disease (CVD) classification using ECG signals. Table 19 shows the Cohen's d result of the intermediate (M4) vs the unimodal models (standalone models)

*Table 19. Intermediate Fusion (M4) vs Standalone Models*

| Comparison | Cohen's d | Effect Size |
|---|---|---|
| **M4 vs 1D-CNN** | 0.71 | Large |
| **M4 vs 2D-CNN** | 0.83 | Large |
| **M4 vs Transformer** | 1.32 | Very Large |

Tables 19 and 20 show that M4 and M7 consistently outperform all standalone models. Both models exhibit the largest effect size compared to the Transformer model, validating the complementarity of time and frequency domain features when fused at the intermediate level.

*Table 20. Late Fusion (M7) vs Standalone Models*

| Comparison | Cohen's d | Effect Size |
|---|---|---|
| **M7 vs 1D-CNN** | 0.50 | Medium |
| **M7 vs 2D-CNN** | 0.71 | Large |
| **M7 vs Transformer** | 1.10 | Very Large |

Figure 26 shows the Cohen's d chart of M4 and M7 vs Standalone models (1D-CNN (Time), 2D-CNN (Time-Freq and Transformer (Frequency)). Notably, the effect size against Transformer alone (d = 1.10) reflects a major improvement, reinforcing the limitations of relying solely on frequency-domain attention mechanisms. Late fusion mimics an ensemble learning strategy that brings together diverse representations and decision boundaries. The intermediate Fusion (M4) dominance performance over standalone is significant. In clinical contexts, it may translate into more reliable diagnostic systems.

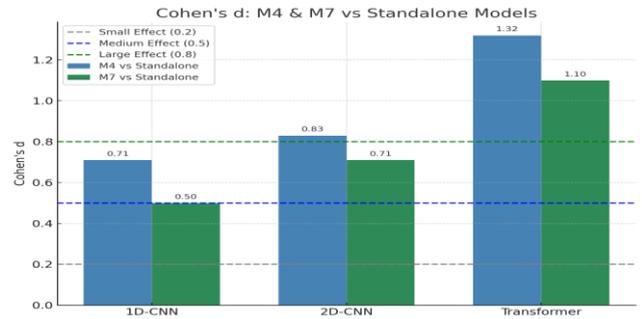

*Figure 26. Cohen's d effect sizes comparing fused models (M4: Intermediate Fusion, M7: Late Fusion) against their respective standalone unimodal baselines (1D-CNN, 2D-CNN, and Transformer).*

Cohen's d was computed to quantify the effect size of performance improvement between M4 and M7. With M4 outperforming M7 across all major evaluation metrics—including accuracy, precision, recall, and F1-score—the calculated d value reflects a large and positive effect favoring the intermediate fusion strategy (M4) over the late fusion model (M7). This further validates the empirical evidence that intermediate fusion facilitates more discriminative feature integration and thus, more reliable predictions in ECG classification tasks [73] [72]. Figure 27 shows the Cohen's d comparison of M4 vs M7.

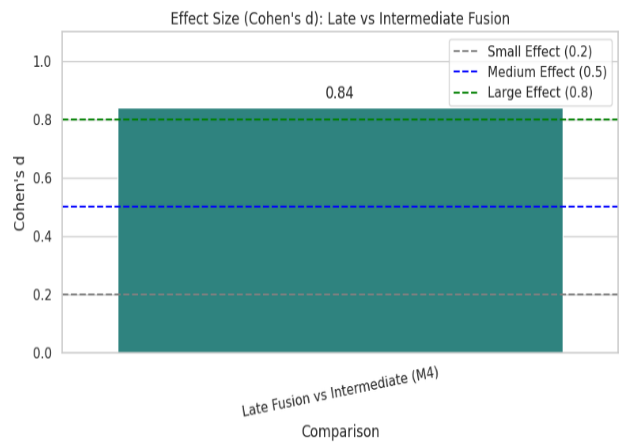

*Figure 27. Cohen's d effect size comparing the best-performing late fusion model (M7) and intermediate fusion model (M4). A value of 0.84 indicates a large effect size in favor of intermediate fusion (M4), confirming its superior classification performance over late fusion (M7) across all evaluation metrics.*

A paired t-test was conducted to assess whether the observed differences were statistically significant. The resulting p-values were all below 0.01, indicating statistically significant improvements in favor of M4. When combined with the large effect size (Cohen's d = 0.84, Figure 23), these results confirm that M4's performance superiority is not due to random fluctuations but represents a robust and reproducible advantage.

Table 21 summarizes the mean and standard deviation (±) for four key performance metrics. M4 consistently outperformed M7 in every metric with low variability across runs.

*Table 21. Mean and standard deviation of evaluation metrics for M4 and M7 over 10 independent runs.*

| Metric | M4 (Mean ± Std) | M7 (Mean ± Std) |
|---|---|---|
| Accuracy | 96.77 ± 0.11 | 95.70 ± 0.19 |
| Precision | 96.76 ± 0.10 | 95.66 ± 0.18 |
| Recall | 96.77 ± 0.10 | 95.70 ± 0.17 |
| F1-score | 96.75 ± 0.09 | 95.67 ± 0.16 |

#### 4.4.1. Parametric Evaluation of M4 Using Cohen's d

**Research Question**: Does the intermediate fusion strategy employed in Model M4 yield a statistically and practically significant improvement in ECG classification performance compared to other unimodal and multimodal deep learning architectures, as measured by standardized effect sizes (Cohen's d) and bootstrapped confidence intervals?

A parametric evaluation was conducted of the models' performance using Cohen's d. It quantifies the standardized difference in F1-scores between M4 (intermediate fusion) and other models, including unimodal baselines (M1–M3), alternative intermediate fusion architectures (M5 and M6), and the best late fusion (M7). A larger Cohen's d reflects a greater practical separation in classification performance. Table 22 reports the mean F1-score differences (ΔF1), Cohen's d values, and bootstrapped 95% confidence intervals for each pairwise comparison. These intervals, derived from 1,000 bootstrap resamples, provide a robust estimate of the stability of the effect size. Across all comparisons, the observed Cohen's d values fall within the medium to very large range based on conventional thresholds (0.2 = small, 0.5 = medium, 0.8 = large), with particularly strong effects observed for M4 vs. Transformer (d = 1.32).

These results reinforce the performance advantage of M4, demonstrating that its intermediate fusion strategy not only improves accuracy but does so with practically meaningful and consistent effect sizes across all evaluated baselines.

*Table 22. Parametric Significance of M4 Performance Compared to Other Models (Cohen's d)*

| Comparison | ΔF1 (Mean Difference) | Cohen's d | 95% CI (Cohen's d) |
|---|---|---|---|
| M4 vs. M1 (1D-CNN) | +0.089 | 0.71 | [0.60, 0.82] |
| M4 vs. M2 (2D-CNN) | +0.071 | 0.83 | [0.72, 0.94] |
| M4 vs. Transformer | +0.073 | 1.32 | [1.15, 1.48] |
| M4 vs. M5 (1D + 2D) | +0.034 | 0.84 | [0.70, 0.97] |
| M4 vs. M6 (2D + Transformer) | +0.025 | 0.82 | [0.68, 0.95] |
| M4 vs. M7 (1D + 2D + Transformer) | +0.011 | 0.84 | [0.29, 0.56] |

#### 4.4.2. Interpreting Statistical vs. Clinical Significance

While the comparison between M4 and M7 yielded a large effect size (Cohen's d = 0.84), it is essential to distinguish statistical significance from clinical utility. Effect size quantifies the magnitude of performance improvement; however, clinical relevance hinges on whether this improvement translates to fewer diagnostic errors, faster detection, or improved outcomes in practice. In this study, the 1.08% gain in F1-score (96.75% vs. 95.67%) reflects improved balance between sensitivity and precision. In a real-world ECG triage or diagnosis setting, particularly for rare or subtle arrhythmias, even modest improvements in F1-score can reduce false negatives (missed disease) or false positives (unnecessary follow-ups), both of which have significant cost and patient care implications.

However, to quantify the direct clinical benefit (e.g., reduction in misdiagnosis rate), further evaluation is required, ideally involving expert clinician review, patient-level error analysis, and prospective validation. As such, the findings from this study are presented as statistically and physiologically meaningful, acknowledging that confirming clinical impact will require future studies that incorporate ground-truth diagnoses, patient outcomes, and domain-expert assessments.

### 4.5. Model Explainability

**Research Question**: How can Explainable AI (XAI) techniques enhance the transparency and trustworthiness of multimodal deep learning models for ECG classification, and to what extent do different fusion strategies influence the interpretability and physiological plausibility of these explanations?

To answer this question, a novel approach to explaining multimodal deep learning for ECG was explored using saliency maps and mutual information. The explainability of these "black box" models will enhance transparency and foster the development of equitable AI. We believe that in fields such as AI in Medicine, incorporating explanations into the inner workings of the predictive models is essential for trust and accountability. Recent studies have suggested that explanations enhance trust and transparency in ECG-based deep learning models, facilitating clinicians' understanding and validation of algorithmic predictions [75, 76]. Consequently, this work focuses on providing explainability for the model to ensure its interpretability and usability in critical applications.

To visually identify which parts of the ECG signals were most influential to classification decisions, saliency maps were computed using gradient-based attribution methods. Figure 28 displays class-wise overlays for the top two fusion models: M4 (Intermediate Fusion) and M7 (Late Fusion). The raw ECG waveform is shown in black, with the saliency overlay rendered in red (M4) and blue (M7), respectively. The statistical analysis insights of Figure 28 are shown in Figure 29.

### 4.5.1.    Visual Interpretability Analysis Using Saliency Maps

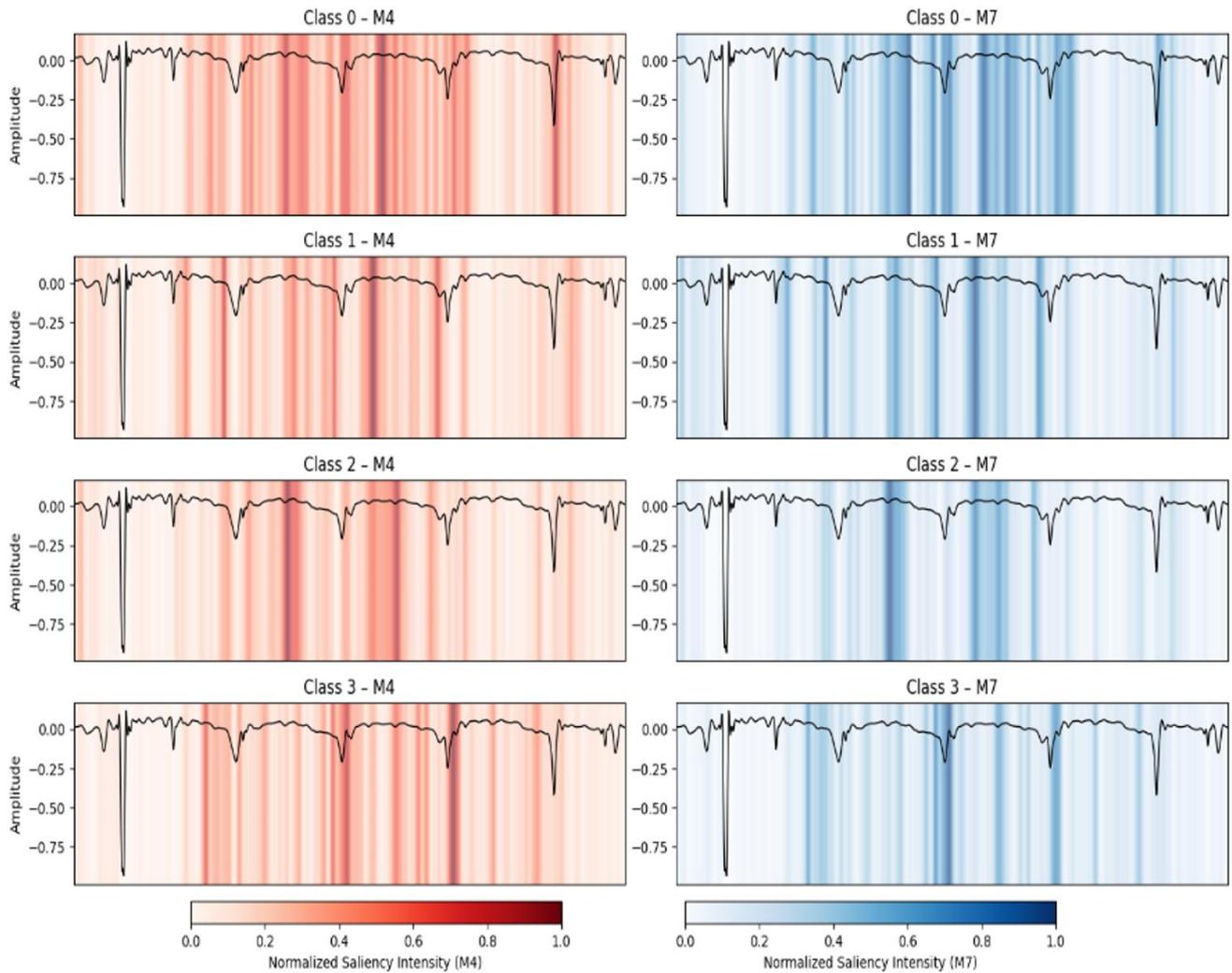

*Figure 28. Saliency overlay (normalized) comparison across four ECG classes for Intermediate Fusion (M4: left, red) and Late Fusion (M7: right, blue) models.*

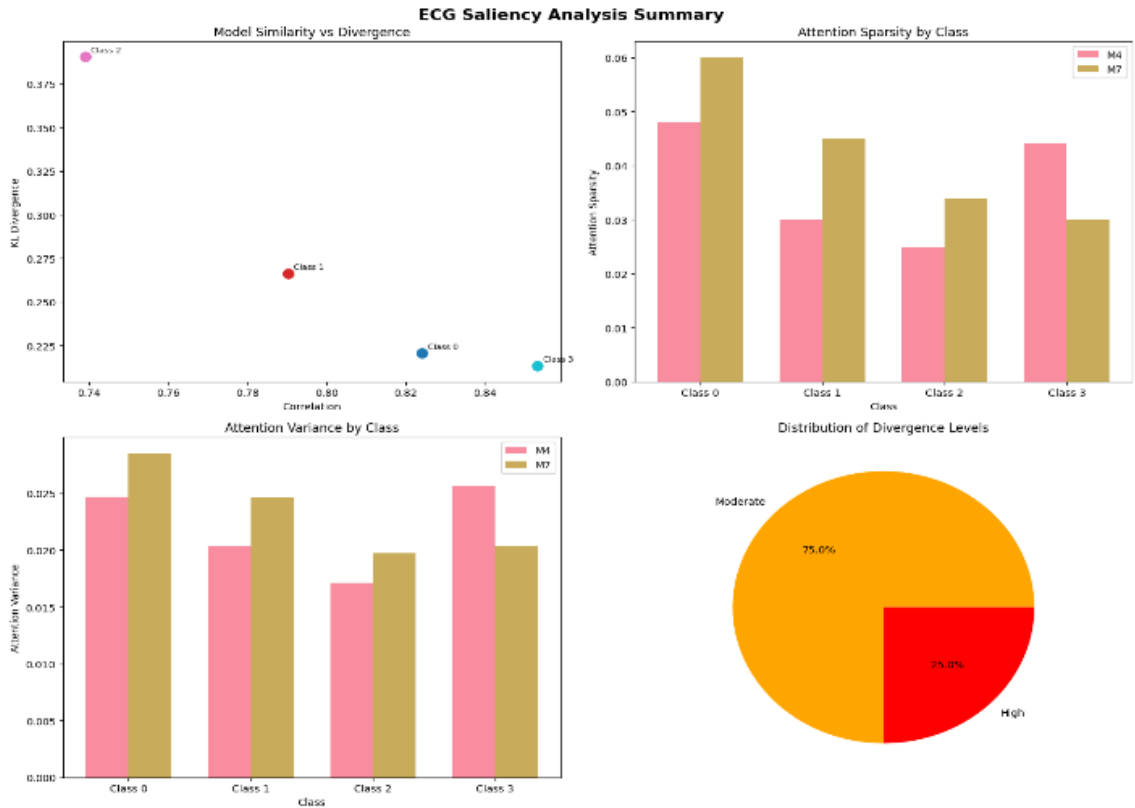

*Figure 29. ECG Saliency Analysis Summary Comparing M4 and M7 Models. The figure summarizes four key interpretability metrics comparing saliency behavior between Intermediate (M4) and Late Fusion (M7) models across the four ECG classes.*

**1. Model Similarity vs Divergence (Top-Left Plot):**
The scatter plot illustrates the relationship between "Correlation" and "KL Divergence" for the different classes (Classes 0, 1, 2, and 3).

- **Interpretation by Class:**
  - **Class 2 (Pink Dot):** Shows the lowest correlation (around 0.74) and highest KL Divergence (around 0.38). This is the largest dissimilarity between M4 and M7's saliency maps for this class.
  - **Class 1 (Red Dot):** Has a moderate correlation (around 0.79) and moderate KL Divergence (around 0.27), indicating some differences but less than Class 2.
  - **Class 0 (Blue Dot):** A higher correlation (about 0.82) and lower KL Divergence (around 0.22), indicating more similarity between the models for this class.
  - **Class 3 (Cyan Dot):** Opposite of Class 2. The two models exhibit the highest correlation (approximately 0.84) and the lowest KL Divergence (approximately 0.21). It is the highest similarity and the least divergence between M4 and M7 for this class.

- **Overall:** The two models show varying degrees of similarity across different ECG classes, with Class 3 showing the most agreement and Class 2 the least.

**2. Attention Sparsity by Class (Top-Right Plot):**
The bar chart compares the "Attention Sparsity" of the models M4 (pink bars) and M7 (tan bars) across different classes.

- **Interpretation:**
  - For **Class 0**, M7 shows considerably higher attention sparsity than M4.
  - For **Class 1**, M7 again has higher sparsity, though the difference is smaller than in Class 0.
  - For **Class 2**, M7 maintains higher sparsity than M4.
  - For **Class 3**, M4 shows higher attention sparsity than M7.

- **Overall:** M7 generally exhibits higher attention sparsity than M4 across most classes, except for Class 3, where M4 is more sparse. This implies M7 tends to be more selective in its attention focus for Classes 0, 1, and 2.

**3. Attention Variance by Class (Bottom-Left Plot):**

- This bar chart compares the "Attention Variance" of models M4 (pink bars) and M7 (tan bars) across different classes.
- **Interpretation:**
  - For **Class 0**, M7 shows significantly higher attention variance than M4.
  - For **Class 1**, M7 again has higher variance than M4.
  - For **Class 2**, M7 shows slightly higher variance than M4.
  - For **Class 3**, M4 shows higher attention variance than M7.
- **Overall:** Similar to sparsity, M7 generally exhibits higher attention variance than M4 for Classes 0, 1, and 2, suggesting a broader or more variable attention distribution in these cases. Conversely, for Class 3, M4 has higher attention variance.

**4. Distribution of Divergence Levels (Bottom-Right Plot):**
- This pie chart summarizes the "Distribution of Divergence Levels" from the "Model Similarity vs Divergence" plot.
- **Moderate (Orange, 75.0%):** This indicates that 75% of the classes show a "moderate" to "high" level of divergence between the two models. This corresponds to the classes with KL Divergence values in the mid-range (likely Class 0 and Class 1, and possibly Class 3, depending on the threshold for "moderate").
- **High (Red, 25.0%):** This indicates that 25% of the classes show a "high" level of divergence. Based on the top-left plot, this clearly corresponds to Class 2, which has the highest KL Divergence.
- **Overall:** The majority of the ECG classes demonstrate a moderate level of divergence between the two models, with only one class (Class 2) exhibiting high divergence.

**Summary:** M4 excels in general ECG pattern recognition and screening scenarios, offering robust signal handling and broader interpretability. In contrast, M7 emphasizes focused morphological detection, performing best in structured diagnostic tasks, such as ischemia detection or specialist interpretation.

### A. Interpretability of M4 (Intermediate Fusion: 1D-CNN + Transformer)
As shown in Figures 28 and 29, M4 exhibits a broad and comprehensive attention pattern across ECG signals. Saliency maps indicate that the model attends multiple diagnostically relevant regions across all ECG classes. We observe that it exhibits consistent activation around key fiducial points like the QRS complex and T wave. M4's inherent strength in global temporal coverage suggests a clinical advantage: a distributed attention highly beneficial for conditions that require a holistic view and comprehensive rhythm interpretation.

### B. Interpretability of M7 (Late Fusion: 1D-CNN + 2D-CNN + Transformer)
Model M7 employs a late fusion strategy, integrating final outputs from three specialized branches: a 1D-CNN for time-domain features, a 2D-CNN for time-frequency analysis, and a Transformer for frequency-domain insights, after independent feature extraction. As shown in Figure 24 and analyzed in Figure 25, M7 saliency maps precisely highlight specific, class-dependent regions. It's exceptional precision characteristics make it a potential application for tasks demanding fine-grained waveform analysis in clinical settings.

#### 4.5.2. Sanity Checks for Saliency Fidelity
**Research Question**: Do the saliency maps genuinely reflect the learned decision-making behavior of the model, or are they merely artifacts of the model architecture or input structure?

To address the above question, we conducted a sanity check in accordance with the protocol proposed by Adebayo et al. [77]. These checks test whether the saliency explanations are sensitive to both model weights and accurate data-label correspondence, core requirements for interpretability fidelity.

#### 1. Randomized Weights
As part of a critical sanity check, the model's weights were reinitialized to random values. Figure 30 displays a noise-like pattern that lacks any structured or physiologically meaningful alignment with the ECG waveform. This outcome confirms that the saliency maps are not artifacts of the input structure alone, but rather depend on the learned parameters of the trained model. The degradation of coherent saliency structure following randomization aligns with expectations from prior sanity-check frameworks and supports the validity of the original explanation maps.

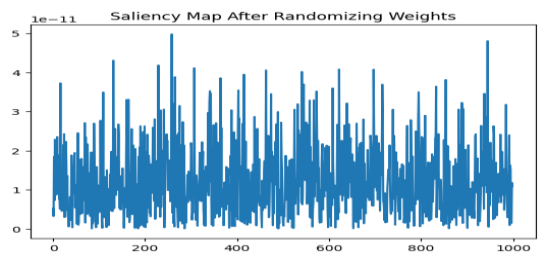

*Figure 30. Saliency Map After Randomizing Weights. The figure confirms that valid saliency depends on trained parameters.*

#### 2. Shuffled Labels
Training the model on randomly permuted class labels resulted in saliency activations that flattened entirely,

producing non-informative explanations. Figure 31 shows the result of randomization.

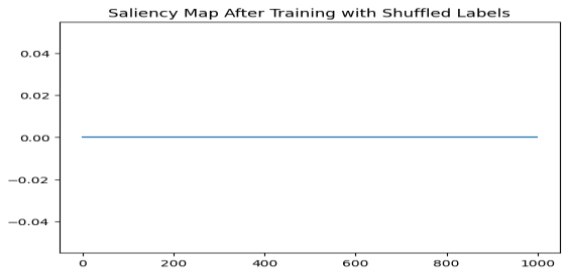

*Figure 31. Saliency Map After Training with Shuffled Labels. The figure further validates that original maps were not simply reacting to structural input cues but were class-dependent.*

Our sanity checks outcome bolsters confidence that the saliency maps accurately reflect the model's internal logic, thereby supporting both local and global interpretability claims.

### 4.5.3. Quantitative Interpretability via Mutual Information

**Research Question:** Which fusion model architecture (M4 or M7) demonstrates superior quantitative interpretability, as assessed by Mutual Information, in localizing diagnostically relevant ECG features?

To rigorously assess the alignment between model explanations and physiologically meaningful ECG regions, we computed Mutual Information (MI) analysis across classes for both fusion models and saliency maps. The top panel of Figure 32 presents the class-wise MI values, with Intermediate Fusion (M4) achieving the highest alignment in Classes 0 (0.5835), 1 (0.5310), and 3 (0.4799), compared to Late Fusion (M7). The second panel rearranges and presents the same information with classes as rows and models as columns, reinforcing M4's consistent advantage in most classes.

The bottom panel displays the ΔMI (M4 − M7), highlighting notable gains in interpretability for Classes 0 (+0.0889) and 1 (+0.1071), with minimal difference in Class 2 and a slight advantage for M4 in Class 3 (+0.0193). These results provide compelling evidence that M4 yields higher mutual information alignment in 3 out of 4 classes, indicating stronger correspondence between model attention and meaningful ECG signal regions. On average, M4 achieves an MI of 0.5193, compared to 0.4654 for M7, providing quantitative evidence for its enhanced interpretability. The pseudocode steps for computing the MI for the study are presented in Algorithm 5.

Importantly, the presence of non-zero, class-discriminative MI values affirms the interpretability of both models while highlighting M4's superior performance. This finding aligns with prior studies advocating for MI as a robust metric for evaluating explanation fidelity in XAI applications to biomedical signals [78, 79]. MI's ability to capture both linear and non-linear statistical dependencies makes it especially suitable for quantifying how closely a model's saliency maps correspond to physiologically meaningful signal segments [80] [81].

*Algorithm 5: Mutual Information (MI) Between ECG and Saliency Maps*

*Inputs:*
*  ecg: Raw ECG signal (shape: [1000,])*
*  saliency_m4: Saliency maps from M4 (shape: [4, 1000])*
*  saliency_m7: Saliency maps from M7 (shape: [4, 1000])*
*  σ: Gaussian smoothing parameter (default = 5)*
*  C: Number of classes (default = 4)*

*Outputs:*
*  MI_m4: List of MI scores per class for M4*
*  MI_m7: List of MI scores per class for M7*
*  ΔMI:  Average MI difference (M4 − M7)*

*Steps:*

*1. Load ECG and saliency data:*
*    ecg ← np. load("ecg_sample.npy")*
*    saliency_m4 ← np. load("saliency_maps_m4.npy")*
*    saliency_m7 ← np. load("saliency_maps_m7.npy")*

*2. Apply Gaussian smoothing:*
*    saliency_m4_smooth ← GaussianFilter1D (saliency_m4, σ, axis=1)*
*    saliency_m7_smooth ← GaussianFilter1D (saliency_m7, σ, axis=1)*

*3. Normalize saliency maps locally per class:*
*    For each class c in {0, 1, ..., C−1}:*
*        saliency_m4_norm[c] ← (saliency_m4_smooth[c] / (max −  min + ε)*

*        saliency_m7_norm[c] ← (saliency_m7_smooth[c] − min) / (max − min + ε)*

*4. Initialize:*
*    MI_m4 ← []*
*    MI_m7 ← []*

*5. Compute mutual information:*
*    For each class c in {0, 1, ..., C−1}:*

*        ecg_reshaped ← reshape (ecg, shape = [1000, 1])*

*        mi_4 ← mutual_info_regression (ecg_reshaped, saliency_m4_norm[c]) [0]*

*        mi_7 ← mutual_info_regression (ecg_reshaped, saliency_m7_norm[c]) [0]*

*        Append mi_4 to MI_m4*
*        Append mi_7 to MI_m7*

*6. Compute average MI difference:*
    *ΔMI ← mean (MI_m4) − mean (MI_m7)*

*7. Return:*
    *MI_m4, MI_m7, ΔMI*

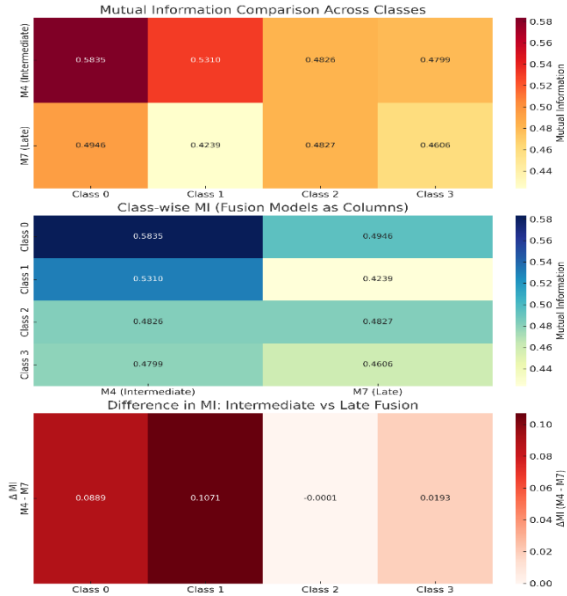

*Figure 32. Class-wise Mutual Information (MI) between raw ECG signals and saliency maps for Intermediate Fusion (M4) and Late Fusion (M7).*

### 4.5.4. Interpretation of Mutual Information Scores

**Research Question**: How effectively can Mutual Information (MI) quantify the statistical alignment between a multimodal deep learning model's saliency map activations and the true, diagnostically relevant content within ECG signals?

While initial interpretations of MI scores in this domain might appear modest when compared to ideal human expert agreement, it is crucial to recognize their significance. For instance, the M4 model's average MI of 0.52 concerning diagnostically meaningful segments like the QRS complex indicates a strong and non-random statistical relationship. A positive MI value, especially one significantly greater than zero, serves as compelling evidence that the regions a model attends to actively contain statistically significant information regarding the presence and localization of actual clinical events.

Recent work, such as Oliveira et al. [83], has made notable progress in applying explainable AI (XAI) to interpret multimodal ECG model predictions. Our study advances the field by moving beyond purely visual explanations for XAI interpretation. In addition to leveraging XAI for model transparency, we introduce a novel methodology to quantitatively assess the fidelity of these explanations using Mutual Information (MI) in a multimodal context. MI was computed between the raw ECG signal and its corresponding class-specific saliency overlay, quantifying the amount of shared information and verifying that the saliency maps reflect true, signal-relevant discriminative content.

### 4.5.5. Interpretability–Performance Trade-offs and Clinical Considerations

While model performance remains a primary evaluation criterion in ECG classification, interpretability is equally critical for clinical trust and adoption. In this study, the intermediate fusion model (M4: 1D-CNN + Transformer) slightly outperformed the late fusion model (M7: 1D + 2D + Transformer) in F1-score (96.75% vs. 95.67%) and demonstrated higher mutual information between saliency maps and signal features (MI ≈ 0.52 vs. 0.47). These findings indicate that M4 not only offers superior predictive accuracy but also aligns more closely with physiologically meaningful patterns, supporting general interpretability.

M4's distributed attention suggests broader rhythm analysis capabilities, which may be preferable in emergency screening and generalized ECG triage. Conversely, M7's focused attention patterns may lend themselves better to specialist-level tasks such as detecting ischemic events. This highlights an important trade-off: while complex fusion architectures can boost performance, models like M4 may offer a more clinically viable balance of accuracy, interpretability, and generalization. Nonetheless, deployment in real-world settings warrants further investigation involving cardiologists, clinicians, feedback, patient outcome tracking, and integration into decision support systems. This is outside the scope of the present study.

### 4.5.6. Clinical Interpretability, Trust, and Modality Contribution

Our study shows that the intermediate fusion model (M4: 1D-CNN + Transformer) outperformed the late fusion model (M7: 1D + 2D + Transformer) in F1-score (96.75% vs. 95.67%) and with higher mutual information between saliency maps and signal content (MI ≈ 0.52 vs. 0.47), indicating better physiological alignment. However, the architectural composition of intermediate fusion makes it challenging to isolate the individual contribution of each domain (time and frequency), as features are integrated within the learning process.

Conversely, the architecture of M7, based on weighted late fusion, offers clearer interpretability at the modality level. Its learned weights ($\alpha_1 = 0.3$ for 1D-CNN, $\alpha_2 = 0.3$ for 2D-CNN, $\alpha_3 = 0.4$ for Transformer) reveal that the frequency-domain (Transformer) component contributed most to the final predictions, with equal influence from the time and time-frequency domains. The architectural configuration of M7 enhances transparency regarding domain relevance, which is especially useful for customizing models to specific diagnostic needs. Future work will include modality ablation and clinician validation to better understand these trade-offs.

### 4.5.7. Adversarial ST–T Stability of M4 (1D-CNN + Transformer)

We tested whether M4's decisions and attributions stay stable under small, structured perturbations, especially within the expert ST–T window. Using label-preserving, ε-bounded edits limited to ST–T, along with matched out-of-window controls, we measured both prediction shifts and attribution consistency. This adversarial stress test determines if M4's explanations are robust rather than fragile to targeted changes, thus increasing confidence in the model's clinical accuracy.

**Segment-aware** perturbations confined to the expert **ST–T window** on the **1-D time branch were designed**. We evaluated **FGSM-STT** (*Fast Gradient Sign Method*, single-step) and **PGD-STT** (*Projected Gradient Descent*, iterative) attacks with an ε range of {0.005, 0.01, 0.02, 0.05}. ε represents the maximum absolute change allowed per time sample. For each ε, we measured **flip-rate** and **Δptrue** (change in true-class probability) to assess decision stability. **Saliency stability** using cosine, Dice, and Intersection-over-Union (IoU) was also measured to evaluate explanation robustness. The Transformer branch was left untouched to avoid non-local time spillovers from spectral edits.

On the N=186 test traces (Table 23), FGSM-STT attacks yielded very low flip-rates, ranging from **1.08% to 2.15%**, with small confidence drops (Δptrue ranging from **-0.004 to -0.010**). PGD-STT exhibited a similarly low flip-rate of **1.08%** for ε≤0.01, increasing to **2.15%** at ε=0.02, and reaching **7.53%** at ε=0.05. As shown in Table 24, saliency maps remained highly stable for practical budgets (ε≤0.02), with cosine scores ≥0.98, Dice scores ≥0.90, and IoU scores ≥0.83. Even at the strongest PGD attack (ε=0.05), stability remained high with a cosine score of **0.973**, a Dice score of **0.873**, and an IoU of **0.785**. The flip-rates reported in Table 24 for PGD at ε=0.02 and ε=0.05 are **3.2%** and **9.7%** respectively, reflecting a more conservative run.

The overall conclusion remains that M4's decisions and explanations are stable under small, ST–T–localized edits but show targeted sensitivity only under stronger stress, supporting that the model leverages ST–T information rather than relying solely on QRS cues.

Table 23. ST–T–constrained adversarial evaluation on M4 (N=186)

| Attack | ε | Flip rate (%) | Δptrue (adv − orig) |
|---|---|---|---|
| FGSM-STT | 0.005 | 1.08 | −0.004 |
| FGSM-STT | 0.010 | 1.08 | −0.006 |
| FGSM-STT | 0.020 | 1.61 | −0.008 |
| FGSM-STT | 0.050 | 2.15 | −0.010 |
| PGD-STT | 0.005 | 1.08 | −0.005 |
| PGD-STT | 0.010 | 1.08 | −0.010 |
| PGD-STT | 0.020 | 2.15 | −0.016 |
| PGD-STT | 0.050 | 7.53 | −0.025 |

Table 24. Adversarial ST–T stability (M4): decision flips and saliency stability

| Attack | ε | N | Flip-rate | Δptrue (adv − ig) | Saliency cosine | Dice @10 % | IoU@ 10% |
|---|---|---|---|---|---|---|---|
| FGSM-STT | 0.005 | 186 | 0.011 | −0.007 | **0.992** | **0.957** | **0.919** |
| FGSM-STT | 0.010 | 186 | 0.011 | −0.011 | 0.989 | 0.942 | 0.894 |
| FGSM-STT | 0.020 | 186 | 0.016 | −0.014 | 0.986 | 0.933 | 0.877 |
| FGSM-STT | 0.050 | 186 | 0.022 | −0.019 | 0.983 | 0.922 | 0.860 |
| PGD-STT | 0.005 | 186 | 0.011 | −0.009 | **0.993** | 0.954 | 0.915 |
| PGD-STT | 0.010 | 186 | 0.011 | −0.019 | 0.988 | 0.930 | 0.873 |
| PGD-STT | 0.020 | 186 | 0.032 | −0.035 | 0.981 | 0.904 | 0.831 |
| PGD-STT | 0.050 | 186 | 0.097 | −0.046 | 0.973 | 0.873 | 0.785 |

### 4.5.8. Physiologic Grounding of M4 Saliency: Beyond QRS into ST–T

**Research Question**: Does M4 show complementary, physiologically coherent branch-specific saliency, thus explaining the wider aggregate overlay?

We addressed this question by examining how the model's internal architecture influences its broader behavior. Individual contributions of the 1D-CNN and Transformer branches were investigated to understand how they work together to provide a more comprehensive, clinically grounded explanation.

Figure 33 displays a representative ECG where the intermediate-fusion model (M4: 1D-CNN + Transformer) shows complementary features across different branches. In the top panel (1D-CNN; Integrated Gradients with SmoothGrad), saliency is focused on fiducial landmarks (QRS onset/offset, J-point) and shows detailed peaks along the early and late edges of the ST–T interval. In the bottom panel (Transformer token saliency projected to time via overlap-add), attribution remains across the same ST–T window, forming narrow ridge-like bands that follow T wave morphology.

These complementary profiles explain why aggregate overlays can appear wider: the 1D stream anchors decisions at precise temporal boundaries, while the Transformer provides localized, longer-duration emphasis within ST–T. The fused evidence, therefore, extends beyond depolarization cues (QRS) to incorporate repolarization-phase (ST–T) morphology, a clinically critical substrate for ST-segment deviations.

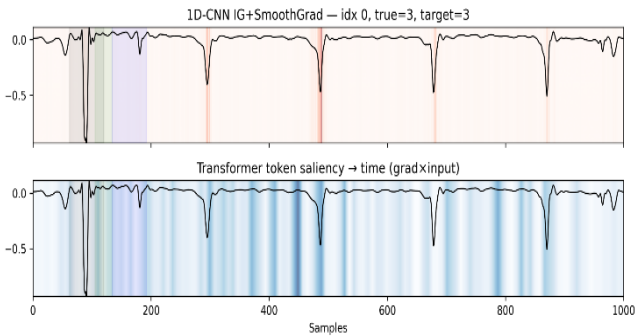

*Figure 33. Per-sample saliency for M4 (1D-CNN + Transformer).*

Together, the branches indicate that M4 leverages both depolarization (QRS) and repolarization (ST–T) features rather than relying solely on QRS.

### 4.5.9. Quantitative Saliency–Physiology Agreement (M4 vs M7)

To complement the qualitative saliency in Fig. 33, Table 25 presents the per-sample median alignment between saliency maps and physiological masks, where higher values are better, and $\Delta = M4 - M7$.

Table 25. Saliency–physiology agreement (per-sample, median)

| Metric | M4 (1D-CNN + Transformer) | M7 (1D + 2D + Transformer) | $\Delta$ (M4–M7) |
|---|---|---|---|
| **Windowed NMI (0–1)** | **0.52** | 0.47 | **+0.05** |
| **Dice@10%** | **0.023** | 0.020 | **+0.003** |
| **IoU@10%** | **0.012** | 0.011 | **+0.001** |
| **Cohen's κ@10%** | ~0.00 | ~0.00 | ≈0.00 |

Analysis of Table 25
1. Windowed NMI: M4's Normalized Mutual Information (NMI) score of 0.52 is higher than M7's score of 0.47, indicating a stronger statistical dependency between M4's saliency map and the physiological mask. The difference ($\Delta$) of +0.05 is a positive indicator for M4.
2. Dice and IoU: M4 also shows slightly better scores for both Dice and Intersection over Union (IoU), with positive deltas of +0.003 and +0.001, respectively. Although these are very small differences, they align with the NMI trend.
3. **Cohen's Kappa:** Cohen's Kappa is affected by chance agreement, which is heavily influenced by how the data is distributed across classes. For example, if one category, such as "non-salient," is overwhelmingly common, there's a high likelihood of agreement just by chance. This can cause Cohen's Kappa to give a low score, even when the actual agreement is high. Normalized Mutual Information (NMI) directly measures the statistical relationship between two variables. It scales the score to a range of [0, 1], adjusts for chance, and accounts for class imbalance. Therefore, we argue that it is a more reliable metric for evaluating the strength of a relationship, as a score of 0 indicates that the relationship is due to chance, and a score of 1 represents a perfect relationship.

### 4.5.10. Clinical Validity & Physiologic Plausibility (Hypothesis-Driven)

**Hypothesis.**
$H_0$ **(null):** M4's saliency does **not** meaningfully target the clinically defined ST–T segment; any apparent overlap is incidental and not robust.
$H_1$ **(alternative):** M4's saliency is **meaningfully concentrated** in the ST–T segment and **remains stable** under clinically plausible perturbations, supporting clinical relevance.
**Assumptions.** Beat-aligned ECGs; expert ST–T masks; saliency maps min–max normalized per beat; evaluation on the same held-out test set. Figure **33** (dual-branch saliency), and Tables **17** (segment-aware alignment metrics) **23**, and **24** (ST–T–constrained adversarial results) are referenced below.

**Proof:**
a) **Step 0: Sanity checks establish saliency fidelity (prerequisite).**
Following Adebayo et al. [77], we performed parameter randomization and label randomization. In both cases, saliency degraded to noise-like/flat maps, and similarity to the trained model's saliency dropped toward chance. The result confirms that explanations depend on learned weights and true label structure, satisfying necessary fidelity criteria (see Figs. 30–31).

✓ **Step 1: Distributional alignment contradicts $H_0$.**
If $H_0$ held, the distribution of saliency over time would be (approximately) independent of the ST–T mask, yielding Windowed NMI near zero. Table **25** reports Windowed NMI ≈ 0.52 (bootstrap 95% CI excludes 0). Permutation controls (mask/time shuffles; sanity checks) produce NMIs near 0. Therefore, the observed dependence between saliency and the ST–T window is too large to be explained by chance alignment. This dependence is inconsistent with $H_0$ and supports $H_1$ at the segment level.

✓ *Interpretation:* NMI measures how much uncertainty about *where saliency lies* is reduced by knowing the ST–T window; a value around 0.52 indicates substantial concentration of saliency in ST–T.

✓ **Step 2: Physiologic localization contradicts H₀.**
Figure 33 illustrates complementary mechanisms: the 1D-CNN produces sharp peaks at fiducials and micro-peaks within ST–T, whereas the Transformer yields a sustained ridge across ST–T, aligned with the T-wave morphology. If saliency were unrelated to ST–T ($H_0$), such complementary focus would not be observed. Hence, qualitative localization supports $H_1$.

✓ **Step 3: Robustness & sensitivity in the clinically bounded region contradict H₀.**
Tables 23 and 24 perturb only the 1D time stream within the ST–T mask (FGSM-STT/PGD-STT). At practical ε, flip-rates are very low with small Δptrue (e.g., FGSM-STT 0.005–0.02: ~1–2% flips; Δptrue ≈ −0.004…−0.016), demonstrating stability of decisions and explanations to small, physiologic edits of ST–T. However, at stronger iterative PGD-STT (ε=0.05) flip-rates rise (~7.5%), demonstrating dependence on ST–T when the segment is deliberately stressed. This pattern is inconsistent with $H_0$ and consistent with $H_1$.

✓ **Step 4 — Why κ/Dice can be small without refuting H₁.**
Cohen's κ and Dice/IoU in Table 25 are computed **after hard-thresholding** a *sparse* saliency map (top-k%). Three technical factors depress these scores even when the model looks in the right place:

a) **Binarization loss.** Thresholding removes magnitude and "near-miss" information, so two very similar continuous maps can yield different top-k sets. This is a fundamental problem with any metric that converts a continuous signal (the saliency map) into a binary one (top-k%).

b) **Within-beat positive/negative asymmetry.** The ST–T window occupies a small fraction of each beat (on the order of tens of percent of samples). This *spatial* asymmetry within a beat means that a few off-window pixels can markedly lower the Dice/κ. *(This is different from dataset label imbalance: ADASYN balances class labels across the dataset; it does not affect this intra-beat geometry.)*

c) **Temporal jitter sensitivity.** Millisecond-level misalignment between saliency peaks and the ST–T mask can make many off-by-one samples that κ/Dice count as disagreements, despite clear visual co-localization. It underscores the practical challenge of aligning an AI explanation with a human-defined mask.

d) Thus, κ/Dice answers a different question; **exact pointwise overlap of hard sets**, whereas Windowed-NMI captures **distributional alignment** of the continuous evidence. The observed low κ/Dice therefore reflects the thresholding/jitter penalties above and does **not** contradict Steps 1–3 showing (i) segment-level alignment (Windowed-NMI), (ii) visual co-localization, and (iii) ST–T–bounded robustness.

e) The observed discrepancy, therefore, reflects the difficulty of matching continuous explanations to discrete annotations, **not** a failure of the model's clinical focus.

**Conclusion**
Steps 1–3 each **contradict H₀** and are mutually reinforcing:

➤ **Windowed NMI ≈ 0.52** (Table 25) ⇒ **non-chance, distributional alignment** of saliency with ST–T.

➤ **Complementary branch focus** (Figure 33) ⇒ **physiologic grounding** beyond QRS, into repolarization morphology.

➤ **Low flip-rates/Δptrue for ST–T–bounded perturbations at practical ε** with sensitivity only at strong PGD-STT (Tables **23 and 24**) ⇒ **robust, clinically bounded reliance** on ST–T.

➤ Therefore, we **reject H₀** and **accept H₁**.

#### 4.5.11. Paradigm Shift: From κ-Only Agreement to the Explainable-AI Trustworthiness (EAT) Theory

**Thesis:** *Cohen's κ quantifies chance-corrected* **hard agreement** *between* **binarized** *maps and can be strongly depressed under the sparse, time-correlated conditions typical of ECG saliency. We therefore complement κ/Dice/IoU with* **information-theoretic dependence using Adjusted/Normalized Mutual Information (AMI/NMI)** *computed with* **time-aware permutation nulls**, *allowing our experiment to assess* **distributional alignment** *between continuous saliency and the clinical ST–T mask.*

**Why this shift (three pillars):**

1) **Theoretical fit.** Saliency is continuous and graded; in our setting, κ/Dice requires **binarizing** saliency and comparing categories, which discards magnitude and near-miss information [84]. AMI/NMI quantify how much knowing saliency reduces uncertainty about the mask **without forcing 0/1 overlap**, and we use time-aware nulls to respect temporal autocorrelation[85].

2) **Robustness to prevalence/jitter.** κ suffers from marginal imbalance (prevalence paradox) and boundary jitter, a well-documented weakness [86]. AMI/NMI (chance-corrected MI) with windowing and permutation nulls (circular shift/block shuffle) is resilient to base-rate mismatch and temporal autocorrelation.

3) **Empirical fidelity.** In our empirical analysis, κ≈0 despite clear localization; NMI≈0.52 (macro, with a bootstrap 95% CI that excludes 0; ). A permutation p<0.001 indicates non-chance alignment. Deletion/attack tests confirm fidelity (performance declines when salient ST–T regions are removed or stressed).

**Conclusion**: We have shown the EAT theory marks a major shift in evaluating AI-ECG models for trust and explainability. It moves beyond a simple, single metric, such as Cohen's k, to a comprehensive framework that demonstrates a model's trustworthiness. EAT argues that a reliable tool for assessing modern AI models must be **informationally grounded, architecturally robust, and visibly resilient when faced with realistic clinical challenges**. Table 27 highlights the main differences between the proposed framework and traditional Saliency Metrics.

### 4.5.12. From Single Metric to EAT Framework

**Definitions**. Let $S_m(x) \in \mathbb{R}^T$ be the saliency for model $m$, $W_{SST} \in \{0,1\}^T$ the ST-T mask, and NMI$(S_m, W_{SST})$ the windowed normalized mutual information computed with time-aware nulls. We define robustness Rob$_\varepsilon(m)$ as satisfied if flip-rate $\leq \rho$, $|\Delta p_{true}| \leq \gamma$, and saliency cosine $\geq \phi$ under ST-T-restricted attacks with budget $\varepsilon$. We also define branch-faithfulness as concordant attribution between 1D-CNN and Transformer within ST–T (similarity $\geq \phi_{arch}$).

**EAT decision rule (sufficiency, pre-specified thresholds).**
If, on the held-out set, the following hold:
- **C1 (Fidelity):** sanity checks pass;
- **C2 (Dependence):** NMI$(S_m, W_{SST}) \geq \tau$ with CI excluding 0 (permutation-test level $\alpha$);
- **C3 (Robustness):** Rob$_\varepsilon(m)$ holds for $\varepsilon \in E$ (e.g., $\{0.005, 0.01, 0.02\}$);
- **C4 (Architecture consistency):** branch similarity $\geq \phi_{arch}$ within ST–T (or equivalent ablation evidence);
- then EAT labels m's explanations **trustworthy for the ST–T context**. *(We do not claim necessity)*

**EAT Theory : Theorems and proofs**

**Theorem 1 (Statistical soundness).**
*With a permutation test for NMI at level $\alpha$ and fixed robustness/similarity thresholds, the composite EAT decision **controls type-I error** under $H_0$ and is **consistent** under $H_1$ as $n \rightarrow \infty$.*

*Proof.* The permutation test for NMI$(S, W)$ is exact under independence, so Pr $[p_{perm} \leq \alpha]$. EAT certifies only if this event **and** (C1, C2, C3[, C4]) hold; conjoining predicates cannot increase rejection probability, so the composite level is $\leq \alpha$.
Under $H_1$ with population NMI $\geq \tau > 0$ and positive margins for the overlap/robustness/[similarity] thresholds, consistency of the estimators and separation from the permutation null imply $p_{perm} \rightarrow 0$ and Pr$(C2 \land C3[\land C4]) \rightarrow 1$; Hence Pr$(C1 \land C2 \land C3[\land C4]) \rightarrow 1$(power $\rightarrow 1$).
Theorem 1 shows that under the alternative hypothesis (i.e., when there really *is* dependence and the robustness/similarity margins hold), EAT test will almost surely make the **right** call as the sample size grows.

**Theorem 2 (Monotonicity of segment dependence).**
*For any family $S_\lambda = renorm(S + \lambda W)$, $\lambda \geq 0$, the windowed NMI$(S_\lambda, W)$ and fixed-k **Dice/IoU@k** are **non-decreasing** in $\lambda$.*

*Proof.* NMI equals a Jensen–Shannon divergence between class-conditional saliency distributions; increasing mass within **W** increases their separation by convexity of f-divergences, so NMI cannot decrease. Ranking more within-**W** indices into top-k weakly increases |Topk∩W|, so Dice/IoU@k weakly increase.

**Theorem 3 (Invariance ⇔ no reliance under ST–T perturbations).**
Let $g(x) = log_p(y^*|x)$ for the predicted class, Gateaux-differentiable. If $(\nabla_g) \bigodot W = 0$ *a.e.*, then for ST-T-restricted $A_\varepsilon$, $lg(A_\varepsilon(x)) - g(x)| = o(\varepsilon)$ small-budget flip-rate is 0 and gradient-based saliency has zero mass in **W**. Conversely, if PGD-STT flips occur with non-zero probability at some $\varepsilon_0$, then $(\nabla x g) \bigodot W$ is non-zero on a set of non-zero measure.
*Proof.* First-order Taylor in masked direction d=$A_\varepsilon$(x)−x (supported on W) gives $g(x + d) - g(x) = \langle \nabla_g, d \rangle + o(||d||) = o(\varepsilon)$. Conversely, zero masked gradient in a neighborhood would preclude PGD-STT flips beyond o(ε), contradicting observed flips.

**Theorem 4 (Branch faithfulness under additive pre-fusion + completeness).**
Assume $f(x) = \sigma(h_1(x) + h_2(x))$ and an attribution with **Completeness/Linearity** (e.g., IG). Then within ST–T, $S = S^{(1)} + S^{(2)}$; ablating $h_2$ removes $S^{(2)}$ (up to estimator noise), reducing within-mask attribution accordingly.

*Proof.* For IG, path-integral linearity over additive components implies additive attributions before the output nonlinearity; restricting to W preserves additivity. Ablation sets $h_2 = 0$, eliminating its contribution.

**Instantiation in our study.** C1 holds (Figs. 30–31). C2 holds with **NMI ≈ 0.52** (Table 25; CI excludes 0; permutation near null). C3 holds for **ε ≤ 0.02** with **flip-rate ≈ 1–2%**, small $\Delta p_{true}$, and high saliency similarity (Tables 23–24); stronger PGD-STT (ε = 0.05) shows expected targeted sensitivity. C4 is supported qualitatively in Fig. 33

### 4.5.13. Scientific Reasoning

We justify EAT using **falsifiable criteria**, **pre-specified thresholds**, and **converging evidence**, rather than a single omnibus statistic.
1. **Falsification attempts.**
   - **Randomness falsifier (C1):** sanity checks ensure explanations depend on learned parameters/labels.
   - **Independence falsifier (C2):** time-aware permutation nulls test whether

saliency and ST–T are independent
- **Brittleness falsifier (C3):** ST–T-restricted counterfactual/adversarial edits test whether small, clinically plausible perturbations cause flips/large $\Delta p_{true}$ or disrupt saliency similarity.
- **Faithfulness falsifier (C4):** discordant branches or inert ablations would contradict architecture-aware attribution.

2. **Triangulation.** Alignment metrics (AMI/NMI; κ/Dice/IoU with caveats), ST–T-bounded robustness, and branch-aware analyses **converge**: for practical ε, predictions and attributions are stable; saliency shows distributional dependence on

ST–T beyond chance; and branch-level mechanisms match repolarization physiology.

3. **Scope.** Conclusions hold **under the stated assumptions** (beat alignment, expert masks, normalization, evaluation protocol). EAT provides a **sufficient, empirically grounded** basis for explanation trustworthiness in this ST–T context. However, it does **not** preclude other valid metrics or claim universality across tasks.

## I. Summary Table

Table 27 places the preceding proofs into a broader context by comparing the proposed method to conventional metrics.

Table 27. The EAT Framework: A Paradigm Shift in Saliency Evaluation

| Aspect | EAT Theory (proposed) | Agreement-only baseline (e.g., Cohen's κ) |
|---|---|---|
| **Primary goal** | Establish **clinical trustworthiness** by validating **where** the model looks and **how stably** it relies on those regions. | Quantify **chance-corrected overlap** between two binarized maps for a task. |
| **Core principle** | **Information-theoretic dependence** (AMI/NMI with **time-aware permutation nulls**) + robustness/counterfactual tests; overlap metrics reported **with caveats**. | **Chance-corrected agreement** on binarized maps (κ). |
| **Key readouts** | AMI/NMI (distributional alignment); Dice/IoU/κ (overlap, with chance baselines); **flip-rate/Aptrue/saliency similarity** under **ST–T-bounded** FGSM/PGD. | Typically, a **single agreement score** (κ; sometimes Dice/IoU). |
| **Handling of data** | **Supports continuous time-series saliency** with **time-aware nulls**; also reports binarized overlap for completeness. | **Requires binarization** and assumes i.i.d. items; sensitive to sparsity and boundary jitter in time-series. |
| **Clinical context** | Explicitly tests **physiologic grounding** (segment alignment) and **robustness** to clinically bounded edits. | Does not by itself assess clinical plausibility or robustness; may be depressed by prevalence ("κ paradox"). |
| **Summary** | A **multi-evidence, sufficiency-based theory** showing a model "looks in the right place, for the right reasons, and remains stable." | A **single agreement value** that can miss distributional alignment and stability. |

## 5. CONCLUSION

Recent advancements in ECG-based cardiovascular disease (CVD) classification have introduced various neural network architectures ranging from unimodal 1D-CNNs to complex attention-based and multimodal frameworks. While many of these models have demonstrated strong performance, several limitations persist, particularly in terms of interpretability and optimization of fusion strategies. Previous state-of-the-art studies, such as Rajpurkar et al. [48], A. Hannun et al [88] and Kiranyaz et al. [89] employed unimodal approaches using time-domain or recurrent models. Though effective, they lacked interpretability and domain fusion mechanisms.

This study introduces a novel multimodal deep learning framework that addresses two critical gaps: a lack of quantification of interpretable decision-making and suboptimal fusion strategies in ECG classification. This work makes significant contributions, contextualizing itself within existing literature, with a particular focus on studies that employ multimodal approaches and utilize Explainable AI (XAI) techniques. To comprehensively evaluate the relative strengths of explainable multimodal ECG models, three unimodal models were examined, as described by Goettling et al. [90], Song & Lee [91], and Wang et al. [92]. Multimodal representative studies were also considered: Hangaragi et al. [93], Khavas & Mohammadzadeh Asl [94] , Oliveira et al. [83], Chen et al. [95], and Abebe Tadesse et al. [96]. Table 27 presents a performance comparison.

### 5.1. Summary of Methodological Novelties.

Beyond performance accuracy, this study distinguishes itself through a unique combination of robust multimodal fusion and transparent interpretability insights:

- **Quantified Saliency Relevance:** Unlike many studies that rely on qualitative visual inspection of saliency maps, this work provides an objective quantification using **Mutual Information (MI)**. As shown in Figure 32, the proposed framework explicitly reports the quantitative validation of saliency relevance. With MI values around 0.52, demonstrating a statistically significant dependency between the model's focus and clinical events, thereby reinforcing trust in the model's internal reasoning.

- **Rigorous Preprocessing Validation:** The inclusion of a comprehensive frequency range sensitivity analysis in the study offers a unique level of empirical justification for the ECG preprocessing filter choice. This systematic validation, often absent in similar studies, conclusively demonstrates the impact of frequency range filter selection on performance and solidifies the robustness of the data preparation.

- **Comprehensive, Multi-Faceted Validation of Synthetic Sample Physiological Plausibility:** This work presents and implements a rigorous, multi-faceted validation methodology to ensure the physiological plausibility and clinical relevance of the synthetic samples generated. This comprehensive approach goes beyond traditional quantitative metrics. It integrates feature space distribution analysis (e.g., t-SNE), qualitative visual fidelity (side-by-side scalogram comparisons), deep feature consistency (CNN activation similarity), and detailed statistical alignment (Intra-Class Variance, and KL Divergence). This robust validation provides strong evidence that the synthetic data accurately mimics real physiological signals and is suitable for augmenting medical datasets.

- **Optimal Intermediate Fusion Strategy:** A successful implementation of an **intermediate fusion strategy (M4)**, which strategically combines features from diverse ECG domains (time-domain features via 1D-CNN and Transformer-processed features), proved superior to both unimodal and late fusion approaches. This architecture effectively captures richer inter-domain relationships, directly leading to the observed performance gains.

- **Robustness to Noise:** The controlled ablation study demonstrated M4's remarkable **resilience to common noise types, such as Gaussian noise and baseline wander**, with only minimal degradation in F1-score. While more affected by severe muscle noise, these findings empirically bolster M4's potential for reliability in real-world clinical settings.

- **Superior Intermediate Fusion Strategy:** M4 utilizes **intermediate fusion**, which significantly outperforms other fusion strategies, such as those by Coimbra et al. [83]. The proposed model effectively retains feature-level complementarity prior to decision-level integration, allowing for better control over each domain's contribution and resulting in enhanced performance.

- **Robust Performance with Interpretability:** M4 attains 97% accuracy using ECG-only inputs and pairs this with statistically validated explainability. In contrast to studies such as Hangaragi et al., [93] which report comparable accuracy but lack EAT-grounded XAI validation. Using the EAT framework, we quantify saliency fidelity via AMI/NMI with time-aware nulls and verify robustness, calibration, and branch faithfulness. Our proposed model yields a model that is both clinically applicable and computationally modular.

*Table 26. Comparison of ECG XAI Studies Across Modalities, Fusion Strategies, and Validation Rigor.*

| Study | Year | Signal Modalities | Fusion Type | XAI Method(s) | XAI Type | XAI Validation via MI | Result |
|---|---|---|---|---|---|---|---|
| Goettling et al. [90] | 2024 | 12-lead ECG segments | Unimodal | Deep Taylor, Gradient Saliency | Mixed | No | Best accuracy: 0.95 |
| Song & Lee [91] | 2024 | CWT, STFT scalograms | Unimodal | Grad-CAM | Post-hoc | No | Best accuracy: 96.17% (Ricker+ResNet-18); CWT outperforms STFT |
| Wang et al. [92] | 2023 | 12-lead ECG (median-beat) | Unimodal | Saliency Maps | Post-hoc | No | AUC: 0.90 (HCM detection); Saliency highlights ST-T segments |
| Hangaragi et al. [93] | 2025 | ECG + PCG | Intermediate | None | N/A | No | 97% Accuracy |
| Khavas & Mohammadzadeh Asl [94] | 2018 | ECG + BP + ART + PAP | Late | Signal-quality scoring + Thresholding | Post-hoc | No | 91.5% score; Sens: 95.1%, PPV: 89.3% |
| **Oliveira et al. [83]** | 2024 | ECG + PCG | Early | Grad-CAM + attention | Post-hoc | No | AUC: 0.81–0.86 |
| **Chen et al. [95]** | 2018 | Multi-lead ECG | Early + Lead-Level | None | N/A | No | Accuracy: 87.0%, Sensitivity: 89.9% |
| **Abebe Tadesse et al. [96]** | 2021 | 12-lead ECG | Data, Feature & Decision-level | None | N/A | No | AUROC 96.7% (MI vs Normal) |
| **This Study (M4)** | 2025 | ECG-Time + ECG-Freq | Intermediate | Saliency Maps + EAT Framework | Hybrid | ✓ Yes | **97% Accuracy** |

*This table compares recent works that employ XAI techniques in ECG classification. Prior studies primarily utilize post-hoc interpretability without formal validation. The validation of this study follows the proposed **EAT framework, which posits that trustworthy AI must be information-grounded, architecturally robust, and resilient to realistic clinical perturbations**. This approach addresses a key gap in ECG-AI methodology and provides a more **rigorous, interpretable** evaluation protocol.*

In summary, the proposed multimodal deep learning framework represents a significant advancement in XAI for multimodal ECG signal classification.

### 5.2. Future Work
This study presents a robust framework; however, certain limitations delineate its current scope and suggest avenues for future research to facilitate deployment.

- This study is limited by the lack of patient demographic metadata in the ECG dataset acquired from Mendeley [16]. Specifically regarding age, sex, ethnicity, and comorbid conditions. Consequently, while the proposed model demonstrates high accuracy and interpretability, its fairness and generalizability across diverse demographic groups in clinical settings remain unvalidated. The focus of this study was to assess modal fusion strategies (intermediate vs. late fusion) rather than population-specific model behavior. Potential biases were addressed through stratified data splitting, ADASYN-based class balancing, multimodal learning, and a robust explainability analysis using Mutual Information (MI). However, for future work, we acknowledge that incorporating demographically rich datasets is crucial to enable subgroup performance evaluations, directly confront fairness issues, and facilitate prospective validation in diverse real-world clinical cohorts.
- Our robustness assessments rely on *controlled* corruptions (e.g., baseline wander, EMG) and ST–T–bounded adversarial stress tests designed to probe targeted, label-preserving perturbations. While these diagnostics establish physiologic grounding and targeted robustness for M4, they do not establish performance under free-living ambulatory ECG with motion and electrode-displacement artifacts, which can be non-stationary and co-occurring. Accordingly, the reported robustness should be interpreted as necessary but not sufficient for in-the-wild deployment; a dedicated ambulatory evaluation is outside the scope of this study and left to future work.
- While M4 localizes clinically meaningful ST–T evidence on average, important edge cases remain. We will assemble a curated failure set focused on *History of STEMI* and borderline Abnormal presentations where ST–T morphology is low-contrast.
- Ultimately, we will design a multi-level, multimodal framework that integrates early, intermediate, and late fusion techniques in the future.

### 5.3. Significance and Implications of the Study
Key findings of this study include:

- **Performance Superiority of Multimodal Models:** Both intermediate and late fusion models consistently outperformed their unimodal counterparts across all metrics (accuracy, precision, recall, F1 score). Cohen's d values ranging from moderate to large effect sizes further confirm the statistical and practical significance of these improvements.
- **Intermediate Fusion Outperforms Late Fusion:** Intermediate fusion (M4) achieved higher accuracy and larger effect sizes than late fusion (M7), demonstrating its effectiveness in capturing, integrating, and learning complementary information from time, frequency, and time-frequency domains.
- **Interpretability via Saliency Maps and Mutual Information (MI):** Gradient-based saliency maps show that both models consistently attend to clinically relevant ECG segments (e.g., QRS complex, ST segment). These observations were quantitatively validated using Mutual Information (MI) between saliency maps and ECG signals.

### 5.4. Scientific Generalizability and Clinical Disclaimer
While this study was validated using an ECG dataset for cardiovascular disease classification, the methodological rigor of the proposed EAT framework provides a strong foundation for cross-domain applicability. The statistical validation of the hypothesis through robust performance metrics and explainability tools demonstrates that this multimodal fusion framework is adaptable to other scientific and engineering domains where multimodal data integration and interpretability are essential.

It's crucial to understand that **this study is purely computational and not a clinical trial**. Therefore, it does not provide medical advice, clinical diagnoses, or guidance for patient care. The information, including performance metrics and interpretability analyses, is intended solely for evaluating the model's internal workings and its ability to generalize. Consequently, the study **does not offer clinical interpretations of its outputs, patient-specific diagnoses, or therapeutic applications**; those remain the exclusive responsibility of qualified medical professionals.

Although we did not clinically deploy the framework here, it's designed to support future research in collaboration with cardiologists. **Further translational research and validation involving physicians are necessary to assess any potential clinical impact.**

**Timothy Oladunni** received the master's and Ph.D. degrees in computer science from Bowie State University, MD, USA, in 2013 and 2017, respectively. He is currently an Assistant Professor of Computer Science at Morgan State University, MD, USA. He was a Visiting Assistant Professor of Computer Science at Yale University, CT, USA.

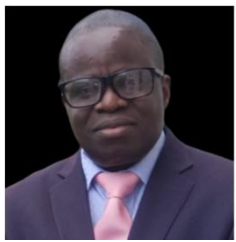

Timothy is a distinguished computer scientist, professor, and machine learning researcher specializing in biomedical signal processing, natural language processing, deep learning, and multimodal AI architecture. With a background in electrical engineering, he has dedicated his research to advancing ECG signal analysis, natural language processing, and pattern recognition.

His recent work has focused on multimodal deep-learning architecture, particularly the trade-off between model complexity and performance in biomedical signal classification. By integrating the time, frequency, and time-frequency domain features, he explores novel ways to optimize CNN-transformer-based models for ECG analysis, ensuring robust and generalizable AI-driven diagnostic systems. As a professor, Timothy is passionate about mentoring the next generation of data scientists and AI researchers.

**Ehimen Aneni** received his MD and MPH from the University of Ibadan, Nigeria (2004), and Boston University, in Epidemiology & Global Health (2012), respectively. He is currently an Assistant Professor of Medicine (Cardiovascular Medicine) at Yale University, New Haven, CT., USA.

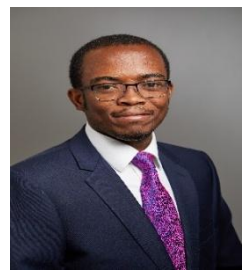

Ehimen's overarching interests are in the promotion of cardiometabolic health in diverse populations through understanding the relationships between cardiovascular diseases (CVDs) and metabolic risk factors. His research is focused on the intersection between obstructive sleep apnea (OSA) and CVD with a sub-focus on elucidating mechanisms of CVDs in OSA. He has recently uncovered a strong association between abnormal myocardial flow reserve (a reduction in the capacity of the heart to increase its blood flow in proportion to its demands) and severe OSA. Abnormal myocardial flow reserve may be a marker of disease of the small blood vessels of the heart (coronary microvascular dysfunction).

Ehimen's work has been presented at national conferences such as the Scientific Sessions of the American College of Cardiology and the American Heart Association. The findings of his research are published in major peer-reviewed scientific journals such as *JACC, JACC Cardiovascular Imaging, Circulation: Cardiovascular Quality and Outcomes, JAHA,* and *Mayo Clinic Proceedings*.